\documentclass[twocolumn]{aastex63}
\usepackage{multirow}
\usepackage{amsmath}
\usepackage[figuresright]{rotating}
\usepackage{threeparttable}
\usepackage{subfigure}
\usepackage{epic,eepic}
\usepackage{graphicx}
\usepackage{longtable}
\usepackage{float}
\usepackage{lineno}
\usepackage{pifont}
\usepackage{hyperref}
\usepackage{gensymb}
 \usepackage{textcomp, gensymb}
 \usepackage{CJK}

\newcommand{\uat}[2]{\href{http://astrothesaurus.org/uat/#2}{#1 (#2)}}

\shortauthors{Han et al.}

\begin{document}

\begin{CJK*}{UTF8}{gbsn}
\title{Stellar X-ray activity and habitability revealed by \emph{ROSAT} sky survey}

%\correspondingauthor{Song Wang}
%\email{songw@bao.ac.cn}

%\author{aaa}
\author{Henggeng Han}%$^\dagger$}
\affil{Key Laboratory of Optical Astronomy, National Astronomical Observatories, Chinese Academy of Sciences, Beijing 100101, People's Republic of China}

\author{Song Wang$^\dagger$}
\affil{Key Laboratory of Optical Astronomy, National Astronomical Observatories, Chinese Academy of Sciences, Beijing 100101, People's Republic of China}
\affil{Institute for Frontiers in Astronomy and Astrophysics, Beijing Normal University, Beijing, 102206, People's Republic of China}
\email{$^\dagger$ Corresponding Author: songw@bao.ac.cn}

\author{Chuanjie Zheng}
\affil{Key Laboratory of Optical Astronomy, National Astronomical Observatories, Chinese Academy of Sciences, Beijing 100101, People's Republic of China}
\affil{School of Astronomy and Space Science, University of Chinese Academy of Sciences, Beijing 100049, People's Republic of China}

\author{Xue Li}
\affil{Key Laboratory of Optical Astronomy, National Astronomical Observatories, Chinese Academy of Sciences, Beijing 100101, People's Republic of China}
\affil{School of Astronomy and Space Science, University of Chinese Academy of Sciences, Beijing 100049, People's Republic of China}

\author{Kai Xiao}
\affil{School of Astronomy and Space Science, University of Chinese Academy of Sciences, Beijing 100049, People's Republic of China}

\author{Jifeng Liu}
\affil{Key Laboratory of Optical Astronomy, National Astronomical Observatories, Chinese Academy of Sciences, Beijing 100101, People's Republic of China}
\affil{School of Astronomy and Space Science, University of Chinese Academy of Sciences, Beijing 100049, People's Republic of China}
\affil{Institute for Frontiers in Astronomy and Astrophysics, Beijing Normal University, Beijing, 102206, People's Republic of China}
\affil{New Cornerstone Science Laboratory, National Astronomical Observatories, Chinese Academy of Sciences, Beijing, 100012, People's Republic of China}

\begin{abstract}
Using the homogeneous X-ray catalog from \emph{ROSAT} observations, we conducted a comprehensive investigation into stellar X-ray activity-rotation relations for both single and binary stars.
Generally, the relation for single stars consists of two distinct regions: a weak decay region, indicating a continued dependence of the magnetic dynamo on stellar rotation rather than a saturation regime with constant activity, and a rapid decay region, where X-ray activity is strongly correlated with the Rossby number.
Detailed analysis reveals more fine structures within the relation:
in the extremely fast rotating regime, a decrease in X-ray activity was observed with increasing rotation rate, referred to as super-saturation, while in the extremely slow rotating region, the relation flattens, mainly due to the scattering of F stars.
This scattering may result from intrinsic variability in stellar activities over one stellar cycle or the presence of different dynamo mechanisms.
Binaries exhibit a similar relation to that of single stars while the limited sample size prevented the identification of fine structures in the relation for binaries.
We calculated the mass loss rates of planetary atmosphere triggered by X-ray emissions from host stars. Our findings indicate that for an Earth-like planet within the stellar habitable zone, it would easily lose its entire primordial H/He envelope (equating to about 1\% of the planetary mass).
\end{abstract}

\section{Introduction} 

The mechanism responsible for coronal heating has long been a widely discussed subject. 
Corresponding theories includes the dissipation of Alfv$\rm{\acute{e}}$n waves \cite[e.g.][]{1947MNRAS.107..211A, 2011ApJ...736....3V} or impulsive energy release due to the dissipation of magnetic stress of bipolar fields, i.e., the \emph{nanoflare} model \citep{1988ApJ...330..474P}. X-ray emission, also named X-ray activity, has been detected on most stars throughout the Hertzsprung-Russell diagram (HRD). For low-mass stars the X-ray activity originates from hot plasma in the corona \citep{1981ApJ...245..163V}, which is strongly related to stellar magnetic fields. 
Therefore, investigating stellar magnetic activity could provide valuable insights into the mechanism behind coronal heating.

Stellar rotation and differential rotation are two key ingredients in stellar dynamo theories \citep[e.g.][]{1955ApJ...122..293P}. Generally, stellar activity proxies, 
including Ca \scriptsize{\uppercase\expandafter{\romannumeral2}} \normalsize H$\&$K lines \citep[e.g.][]{1984ApJ...279..763N, 2018A&A...618A..48M}, $\rm{H\alpha}$ lines \citep[e.g.][]{2017ApJ...834...85N, 2023ApJS..264...12H} and X-ray emissions \cite[e.g.][]{2003A&A...397..147P, 2011ApJ...743...48W}, are strongly correlated to stellar rotation. The consequent stellar activity-rotation relation has been extensively explored from various perspectives, which is expected to shed light on the magnetic dynamos operating on different stars. In general, the stellar activity-rotation relation is divided into two regions, i.e., the saturated region, in which the stellar activity level keeps constant and the decay region, in which activity level is dependent on stellar Rossby number($P_{\rm{rot}} / \tau$) \citep[e.g.][]{1984ApJ...279..763N}. Interestingly, for both dwarfs and giants, which develop different stellar structures, the decay regions of activity-rotation relation obey the same power-law, indicating the development of the same turbulence-related dynamo \citep{2020NatAs...4..658L}.

However, such classical picture may not be universal. In the decay region of the X-ray activity-rotation relation, F-type stars exhibit a possibly different power-law \citep{2019A&A...628A..41P} compared with other type stars.
Similar phenomenon was also found with the $\rm{H\alpha}$ proxy for F stars \citep{2023ApJS..264...12H}. A revising of the X-ray and Ca \scriptsize{\uppercase\expandafter{\romannumeral2}} \normalsize H$\&$K activity-rotation relation shows that a single power-law could not describe the decay region properly \citep{2018A&A...618A..48M}. 
All these studies suggest that different types of stars may develop different dynamos, which could be revealed by the different slopes in the decay region. 

Another hot topic regarding on stellar activity is its influence on planetary habitability. The continuous habitable zone (CHZ) is the region in which liquid water could be maintained on the surface of a planet \citep[e.g.][]{2013ApJ...765..131K}. However, planets in CHZ may still be inhabitable due to strong high energy radiation, including UV and X-ray emissions from their host stars. High energy radiation of host stars plays an important role in stripping planetary atmosphere \citep[e.g.][]{2009A&A...506..399L}, while the composition of planetary atmosphere determines the temperature and pressure of a planet and thus the capability of generating life \citep{2020PNAS..11718264K, 2021AJ....161..213S}. Hence, it is necessary to study the potential impacts of stellar activities on the habitability of planets. 

In this work, we constructed a homogeneous X-ray sample with stellar parameters using \emph{ROSAT} 2RXS catalog together with LAMOST DR10, APOGEE DR16 and Gaia eDR3. Combined with various catalogs of photometric sky surveys, we investigated statistic properties of stellar X-ray activity and its influence on stellar habitable zone. Structure of this paper is as follows. 
In Section 2 we described our sample construction, target classification and cleaning processes.
In Section 3 we discussed the X-ray activity distributions and the activity-rotation relations in detail. 
A discussion on the Ultraluminous sample is given in Section 4. 
In Section 5 we studied the impacts of X-ray emission on planetary atmosphere. 
In Section 6 we presented a brief summery.

\section{Targets selection and Classification}
\subsection{Targets selection}
\label{target.sec}

\begin{figure}
\centering
\includegraphics[width=0.47\textwidth]{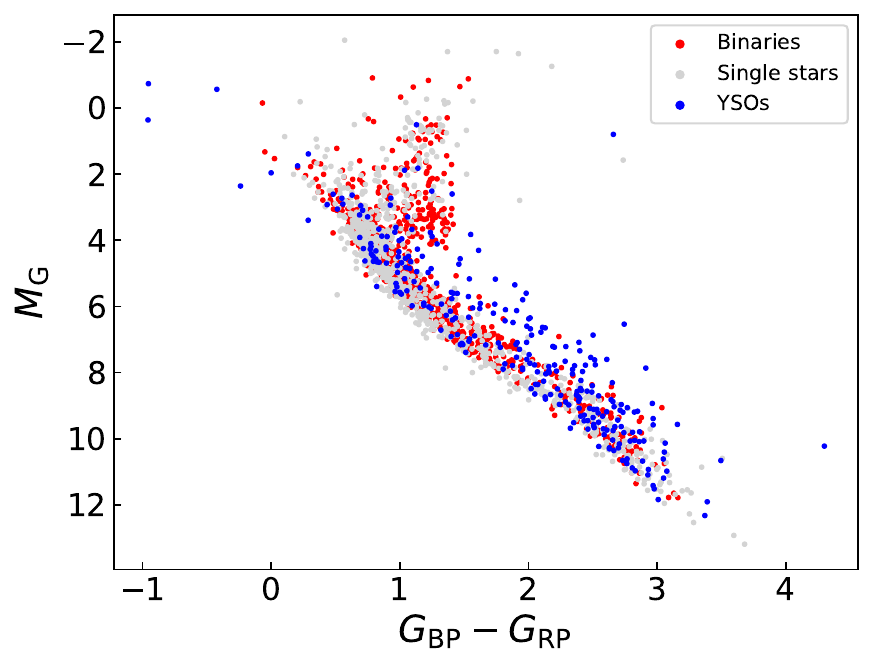}
\caption{Colour-magnitude diagram of our sample. Single stars, binaries and YSOs are shown with grey, red and blue points, respectively.}
\label{HRD.fig}
\end{figure}

The ROentgen SATellite (\emph{ROSAT}) was designed to carry out all sky X-ray survey with high sensitivity in energy band of 0.1--2.4 keV \citep{1982AdSpR...2d.241T}. Consisting of the bright source catalogue and the faint source catalogue, the \emph{ROSAT} 1RXS catalogue contains roughly 127,730 sources
\citep{1999A&A...349..389V}. 
%\citep{1996IAUC.6420....2V, 1999A&A...349..389V, 2000IAUC.7432....3V}. 
Later on, the \emph{ROSAT} point source catalogue underwent a renewal and was named as the Second \emph{ROSAT} All-Sky Survey Point Source Catalog(RASS2RXS), which contains 135,000 X-ray sources \citep{2016A&A...588A.103B}. 
In this study, we used the \emph{ROSAT} 2RXS catalogue.
%For the present study, the selected dataset is the \emph{ROSAT} 2RXS catalogue.

\begin{table*}
  \centering
  %\begin{center}
    \caption{Stellar parameters of each target in our sample.}
    \label{tab:table1}
    \begin{tabular}{ccccccccc}
    %\begin{tabular}{p{3cm}<{\centering}p{1.6cm}<{\centering}p{1.25cm}<{\centering}p{1.25cm}<{\centering}p{1.6cm}<{\centering}p{1.25cm}<{\centering}p{1.25cm}<{\centering}p{1.25cm}<{\centering}p{1.25cm}<{\centering}p{1.25cm}<{\centering}p{1.25cm}<{\centering}p{1.25cm}<{\centering}p{1.25cm}<{\centering}}
      \toprule % <-- Toprule here
       Gaia ID & 
       R.A. &
       Decl. &
       $T_{\rm eff}$ &
       $\rm{log\emph{g}}$ &
       [Fe/H] &
       Distance &
       Radius &
       Type\\
       & (degree) & (degree) & (K) & (dex) & (dex) & (pc) & ($R_{\odot}$) &   \\
       (1) & (2) & (3) & (4) & (5) & (6) & (7) & (8) & (9)\\
      \hline \\ % <-- Midrule here
158364034415872 & 45.540909 & 0.77053 & 3636$\pm$81 & 5.03$\pm$0.16 & -0.32$\pm$0.21 & 371$\pm$12 & 0.49$\pm$0.03 & Dwarf\\
7798908990819968 & 46.445812 & 7.760454 & 6605$\pm$18 & 4.17$\pm$0.02 & 0.1$\pm$0.01 & 284$\pm$13 & 2.08$\pm$0.16 & Dwarf\\
9124679495764864 & 50.91339 & 5.687451 & 3131$\pm$105 & 4.44$\pm$0.24 & -0.19$\pm$0.32 & 37$\pm$1 & 0.49$\pm$0.02 & YSO\\
9770058461712000 & 52.784938 & 7.223575 & 5039$\pm$57 & 4.38$\pm$0.08 & 0.04$\pm$0.04 & 100$\pm$1 & 1.05$\pm$0.08 & Binary\\
9964981257351936 & 51.278631 & 7.393461 & 6052$\pm$127 & 4.2$\pm$0.08 & 0.05$\pm$0.01 & 127$\pm$1 & 1.31$\pm$0.03 & Binary\\
17644790803959552 & 51.831128 & 13.365846 & 4965$\pm$33 & 4.42$\pm$0.05 & -0.05$\pm$0.03 & 200$\pm$1 & 1.08$\pm$0.04 & Binary\\
18211043587721088 & 40.357975 & 5.9884 & 4073$\pm$58 & 4.14$\pm$0.07 & -0.33$\pm$0.03 & 44$\pm$1 & 1.0$\pm$0.02 & Dwarf\\
20896222781456768 & 43.810399 & 9.111804 & 6062$\pm$20 & 4.36$\pm$0.02 & 0.15$\pm$0.01 & 146$\pm$1 & 1.27$\pm$0.08 & Dwarf\\
24331990460013568 & 36.354505 & 10.464996 & 5384$\pm$56 & 4.36$\pm$0.08 & 0.42$\pm$0.05 & 371$\pm$2 & 0.94$\pm$0.07 & Binary\\
27487726270657536 & 45.346629 & 11.396368 & 3930$\pm$66 & 5.03$\pm$0.14 & -0.47$\pm$0.2 & 65$\pm$1 & 1.09$\pm$0.13 & YSO\\
29835488537920768 & 47.791111 & 14.403983 & 6313$\pm$13 & 4.11$\pm$0.02 & -0.1$\pm$0.01 & 327$\pm$2 & 1.65$\pm$0.01 & Dwarf\\

... & ... & ... & ... & ... & ... & ... & ... & ... \\
      \hline \\% <-- Bottomrule here
    \end{tabular}\\
  %\end{center}
%\end{table*}
\footnotesize{(1) Gaia ID: Gaia eDR3 source ID. (2) R.A.: right ascension (3) Decl.: declination. (4) $T_{\rm{eff}}$: effective temperature. (5) log$g$: surface gravity. (6) [Fe/H]: metallicity. (7) Distance: Gaia distance. (8) Radius : stellar radius. (9) Type: stellar types}
\end{table*}
Recent spectroscopic surveys have provided tens of millions of stellar spectra, which can be used to derive stellar atmospheric parameters. 
For example, the LAMOST DR10 low-resolution catalogue contains over 10 millions of spectra of A-, F-, G-, K- and M-type stars \citep{2012RAA....12.1197C, 2015RAA....15.1095L},
while the Apache Point Observatory Galactic Evolution Experiment (APOGEE)-2 project has released 473,307 spectra of 437,445 stars \citep{2020AJ....160..120J}.

Due to the poor positional accuracy of the \emph{ROSAT} 2RXS catalog, 
we first cross-matched the \emph{ROSAT} 2RXS catalogue with LAMOST DR10 and APOGEE DR16 datasets using TOPCAT\footnote{http://www.star.bris.ac.uk/$\%$7Embt/topcat} with a radius of 10". 
The stellar parameters were extracted from the spectra with the highest signal to noise ratio.
Then the sample was cross-matched with Gaia eDR3 \citep{2021A&A...649A...1G} and its distance catalog \citep{2021AJ....161..147B} , with a radius of 3", in order to derive stellar parallaxes, distances and Gaia magnitudes.
To ensure the reliability of distances, only targets within 5 kpc with a relative parallax error (calculated as the parallax error divided by parallax) smaller than 0.2 were remained. 
We derived an initial sample containing 2085 targets with X-ray emission and stellar parameters.

When cross-matching an X-ray catalog with optical catalogs, the standard approach is to identify the nearest optical target (to the X-ray spatial position) as the counterpart of a X-ray source. Due to the poor position uncertainty of \emph{ROSAT} sources, it is necessary to verify the accuracy of the cross-match process and the false match rate of the optical counterparts.

Considering the high positional accuracy of {\it Chandra} observations, the false match rate between {\it Chandra} and optical catalogs using the nearest counterparts could be low.
Here we used the {\it Chandra} sources to roughly estimate the false match rate of this study.
we cross-matched the \emph{ROSAT} 2RXS catalog and the \emph{Chandra} point source catalog \citep{2016ApJS..224...40W}. For each \emph{ROSAT} target, all $Chandra$ counterparts within a 10" radius were picked out. Two selection criteria were applied: (1) selecting the nearest $Chandra$ targets and (2) selecting the brightest $Chandra$ targets. 
Here we assumed the brightest $Chandra$ target is the true counterpart.
This process returned 854 matches, with 775 being identical sources (i.e., the nearest target is also the brightest). Consequently, the false match rate is approximately 9$\%$ ($=$79$/$854). Furthermore, when these two catalogs were cross-matched using radii of 20" and 30", the false match rates are about 11$\%$ ($=$183$/$1613) and 14$\%$ ($=$302$/$2086), respectively. 
Therefore, in our sample, approximately $10\%$ of the optical counterparts to \emph{ROSAT} sources may be incorrectly identified.

\subsection{Target classification and cleaning}
\label{class.sec}

\begin{figure*}
\includegraphics[width=0.5\textwidth]{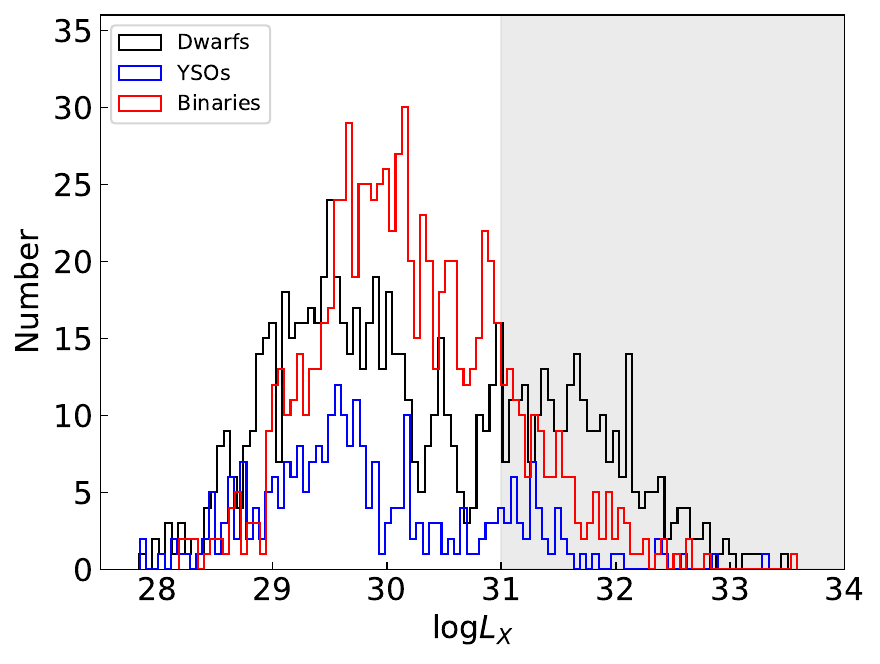}
\includegraphics[width=0.49\textwidth]{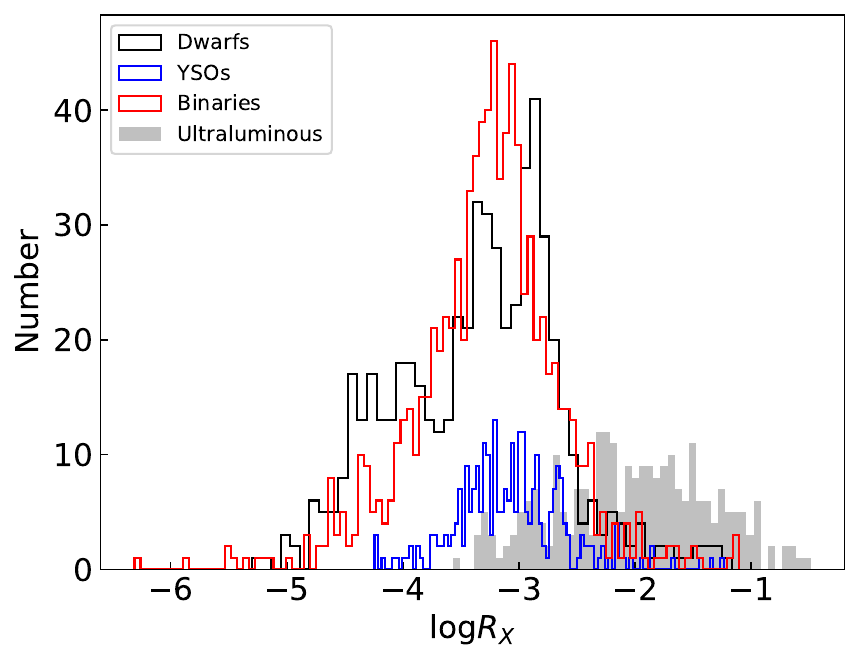}
\caption{Panel (a): Histograms of $L_X$ of dwarfs, giants, YSOs and binaries. Shaded area represents region in which X-ray luminosity is higher than $10^{31}$ erg/s. Panel (b): Histograms of $R_X$ of dwarfs, giants, binaries and Ultraluminous sources.}
\label{Rx_Lx.fig}
\end{figure*}

After the sample construction, the most important process is target classification and cleaning. 
To determine the variable types in our sample, we first cross-matched it with various photometric catalogs, including: (1) periodic variable stars of the Catalina surveys \citep[CSS;][]{2014ApJS..213....9D}; (2) the ASAS-SN catalog of variable stars \citep[e.g.][]{2020MNRAS.491...13J, 2022PASP..134b4201C}; (3) the ZTF variable star catalog \citep[][]{2020ApJS..249...18C}; (4) the WISE variable star catalog \citep{2018ApJS..237...28C}; (5) rotating variables from the \emph{Kepler} and K2 mission \citep{2014ApJS..211...24M, 2013MNRAS.432.1203M, 2019ApJS..244...21S, 2021ApJS..255...17S} and the third version of \emph{Kepler} eclipsing binary catalog\footnote{http://keplerebs.villanova.edu}; (6) the OLGE eclipsing and ellipsoidal binary catalog \citep{2016AcA....66..405S}; (7) the general catalog of variable stars \citep[GCVS;][]{2017ARep...61...80S}; (8) the Gaia variable sources catalog \citep{2023A&A...674A..14R}. (9) two TESS eclipsing binary catalogs \citep{2022RNAAS...6...96H, 2022ApJS..258...16P}. For stars that are identified by multiple catalogs, we graded them according to photometric quality (i.e., \emph{Kepler}, K2, TESS, ZTF, ASAS-SN, CSS, WISE, OLGE, AASVO, GCVS) and used this order to select the first priority classification. 
To further identify possible binaries in our sample, we collected radial velocity (RV) data from LAMOST and APOGEE catalogs. Targets exhibiting RV variations exceeding 20 km/s were considered to be spectroscopic binaries. 
Meanwhile, some double-lined spectroscopic binary (SB2) catalogs have been given from LAMOST spectra \citep{2021ApJS..256...31L, 2022MNRAS.517..356K, 2022ApJS..258...26Z} and APOGEE spectra \citep{2021AJ....162..184K}.
In addition, astrometric binaries were selected based on the catalog of \cite{2021MNRAS.506.2269E}. 

Young stellar objects (YSOs) usually exhibit strong X-ray emission \citep{2003A&A...403..187F} which is unrelated to stellar magnetic dynamo \citep{2003ApJ...582..398F, 2005ApJS..160..390P}.
To classify YSOs from our sample, we cross-matched our dataset with YSO catalogs  \citep{2016MNRAS.458.3479M, 2019MNRAS.487.2522M, 2023A&A...674A..21M} using a radius of 3". 
Among the potential YSOs identified by \cite{2019MNRAS.487.2522M}, only those with a reliability exceeding 90$\%$ were selected. Meanwhile, YSOs that are flagged by SIMBAD were used as supplement. 

Finally, we removed some contamination from our sample. By cross-matching with SIMBAD, objects flagged with Galaxy, QSO, symbiotic stars, high mass X-ray binaries, low mass X-ray binaries and cataclysmic variables were removed.
Binary containing white dwarfs were further excluded based on catalogs from \cite{2018MNRAS.477.4641R, 2020ApJ...905...38R} and \cite{2021MNRAS.506.2269E}.
We also checked the DSS image of each target to avoid the contamination from nebulae and star formation regions.

Figure \ref{HRD.fig} shows the colour-magnitude diagram of the final sample, including 821 dwarfs, 73 giants, 872 binaries and 276 YSOs. In this work, single stars with log$g$ higher than 3.5 were classified as dwarfs and giants are those with log$g$ smaller than 3.5.
Stellar parameters of the final sample are listed in Table \ref{tab:table1}. 

\subsection{Comparison with previous catalogs}
\label{compar.sec}

Based on a Bayesian method, \cite{2018MNRAS.473.4937S} presented a catalog of ALLWISE and Gaia counterparts of \emph{ROSAT} 2RXS targets with Galactic latitudes of $|b| > 15^{\circ}$, offering us an ideal catalog to cross-check our sample. Out of 1589 common X-ray sources, 1440 have the same optical counterparts, leading to a coincidence rate of $\approx$90\%. 
This further confirms the acceptability of the optical counterparts identification in our sample.

\section{X-ray activities and stellar rotation}

\subsection{\emph{ROSAT} X-ray fluxes}

Based on the Interactive Multi-Mission Simulator (PIMMS)\footnote{https://heasarc.gsfc.nasa.gov/docs/software/tools/pimms.html}, \emph{ROSAT} count rate was converted into unabsorbed X-ray flux using the Astrophysical Plasma Emission Code (APEC) model. The plasma temperature was set to be log$T$ = 6.5, considering that most of the targets exhibit plasma temperature between log$T$ = 6 and log$T$ = 7 \citep{2020ApJ...902..114W}.
The metallicity was set to be half of the solar metallitiy.
The galactic extinction was calculated as follows \citep{2017MNRAS.471.3494Z},
\begin{equation}
N_{\rm{H}} = 2.19 \times 10^{21} \ A_V = 6.79 \times 10^{21} \ E(B-V).
\end{equation}
The interstellar reddening $E(B-V)$ was obtained using the three-dimensional dust map \citep{2018JOSS....3..695M} constructed by photometry from Pan-STARRS1 and 2MASS together with Gaia parallaxes \citep{2019ApJ...887...93G}. 
For 484 sources without extinction estimation from the Pan-STARRS1 dust map, we used the SFD dust map \citep{1998ApJ...500..525S} with $E(B-V) =0.884 \times E(B-V)_{\rm SFD}$ as a complement.

\begin{figure*}
%\centering
\subfigure[]{
\includegraphics[width=0.5\textwidth]{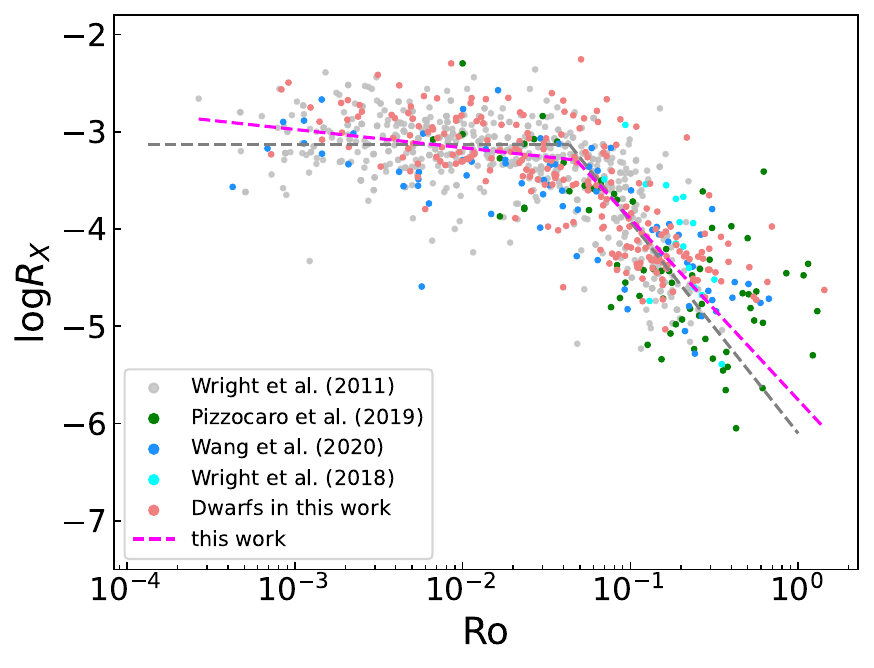}}
%\label{HRD.fig}
\subfigure[]{
\includegraphics[width=0.5\textwidth]{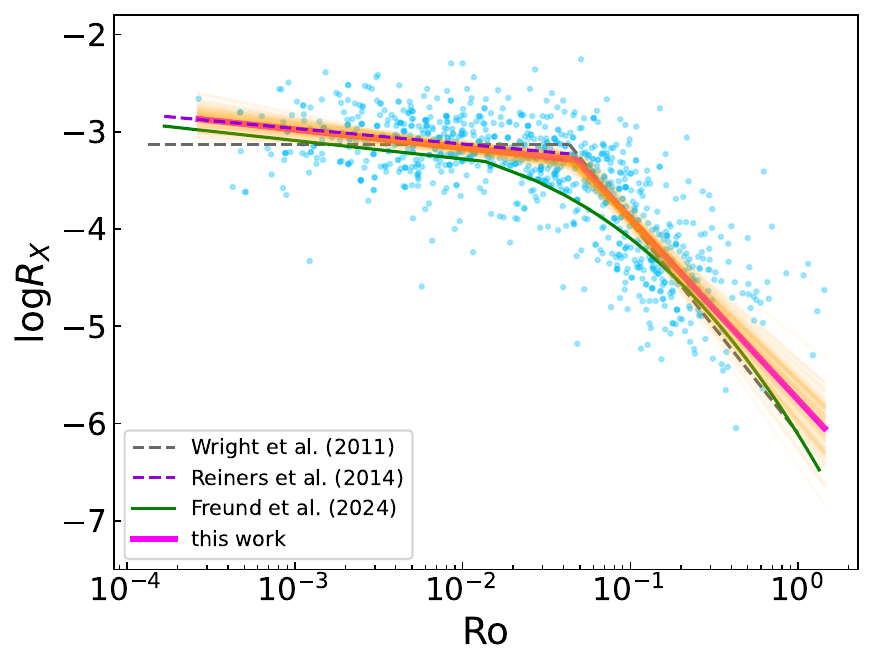}}
%\label{$R_X$.fig}
\caption{Panel (a): Activity-rotation relation of single stars. The grey, green, blue, cyan, and lightcoral dots are from \cite{2011ApJ...743...48W},
\cite{2019A&A...628A..41P},
\cite{2020ApJ...902..114W}, \cite{2018MNRAS.479.2351W}, and this study, respectively. Ro that were empirically estimated are shifted by a factor of Ro$/$3.
The magenta lines are our best-fit results.
Panel (b): Comparisons of activity-rotation relations from different studies. The shaded region shows 100 models extracted randomly from the posterior probability distribution.}
\label{relation}
\end{figure*}

\subsection{X-ray activity}
The X-ray activity level is defined as
\begin{equation}
R_X = \frac{f_{X, \rm{sur}}}{f_{\rm{bol}}} = \frac{f_{X, \rm{obs}}\times(\frac{d}{R_{\star}})^{2}}{\sigma T_{\rm{eff}}^{4}},
\end{equation}
where $f_{X, \rm{sur}}$ is the unabsorbed X-ray flux at the surface of stars and $T_{\rm eff}$ is the stellar effective temperatures. 
The stellar radius was calculated as 
$R_{\rm{*}} =\sqrt{L_{\rm{bol}}/(4\pi \sigma T_{\rm{eff}}^4)}$, with
the bolometric luminosity defined as
$L_{\rm{bol}} = 10^{-0.4\times(m_{\lambda}-5\rm{log}_{10}\it{d}+5-\it{A}_{\lambda}+\it{BC}_{\lambda}-\it{M}_{\odot})}L_{\odot}$.
Here we used the apparent magnitudes in six bands including $G$, $BP$, $RP$, $J$, $H$, and $K_{\rm S}$.
$m_{\lambda}$ is stellar apparent magnitude, and $d$ is the Gaia eDR3 distance \citep{2021AJ....161..147B}. 
$BC_{\lambda}$ is bolometric correction for each band calculated using the \emph{isochrones} python package \citep{2015ascl.soft03010M}, based on stellar effective temperatures, log$g$ and metallicity. 
$M_{\odot}$ and $L_{\odot}$ are solar bolometric magnitude and luminosity, respectively. 
The extinction $A_{\lambda}$ was calculated as $A_{\lambda}=R_{\lambda} E(B-V)$, with $R_{\lambda}$ being the extinction coefficient estimated following \cite{1999PASP..111...63F}.
The mean value and standard deviation of the radii calculated from six bands were used as the final radius and its error, respectively. 
Furthermore, the X-ray luminosity $L_X$ was calculated following $L_X=4\pi f_{X,\rm obs} d^2$. Errors of $R_X$ were derived using a random sampling of $R_X$ based on the errors of $f_{X,\rm{obs}}$, d, $R_{\star}$ and $T_{\rm{eff}}$. Comparison of $R_X$ of common targets between this work and \cite{2011ApJ...743...48W} are given in Figure \ref{cross_wright.fig}, which shows good agreement. 

Figure \ref{Rx_Lx.fig} (Panel a) shows the distributions of $L_X$ for dwarfs, binaries, and YSOs.
One notable feature is the presence of some sources with $L_X > 10^{31}$ erg/s.
Generally, the X-ray luminosity of single dwarfs ranges from $10^{27}$ to $10^{30}$ erg/s \citep{2003A&A...397..147P}.
YSOs can exhibit enhanced X-ray emission with X-ray luminosities about 10 to $10^{4}$ times higher than that of main-sequence stars \citep{1981ApJ...248L..35F, 2002ApJ...574..258F}; in our sample,
most YSOs exhibit a luminosity range of about $10^{29}$ to $10^{31}$ erg/s, suggesting that the majority of YSOs are of class II/III types.

Binaries also present higher luminosities than dwarfs. For example, RS Canum Venaticorum(RS CVn) variables can reach X-ray luminosities ranging from $10^{30}$ to $10^{32}$ erg/s \citep{2010A&A...523A..92M}.
Therefore, it's quite unusual that many single main-sequence stars have $L_X > 10^{31}$ erg/s, leading us to suspect that they may not be of stellar origin.
We categorized them as the ``Ultraluminous" population and excluded them from the dwarf samples. Their nature will be discussed in Section \ref{ultraluminous.sec}.
Figure \ref{Rx_Lx.fig} (Panel b) shows the distributions of $R_X$ for dwarfs, binaries, YSOs, and Ultraluminous sources.
Note that most Ultraluminous sources have $R_X$ values higher than $-$2.5, further evidencing that they are not stars.
Most YSOs exhibit high $R_X$ values around $-$3, while dwarfs and binaries show much wider $R_X$ distributions ranging from $-$5 to $-$2.

\begin{figure*}
\centering
\subfigure[]{
\includegraphics[width=0.47\textwidth]{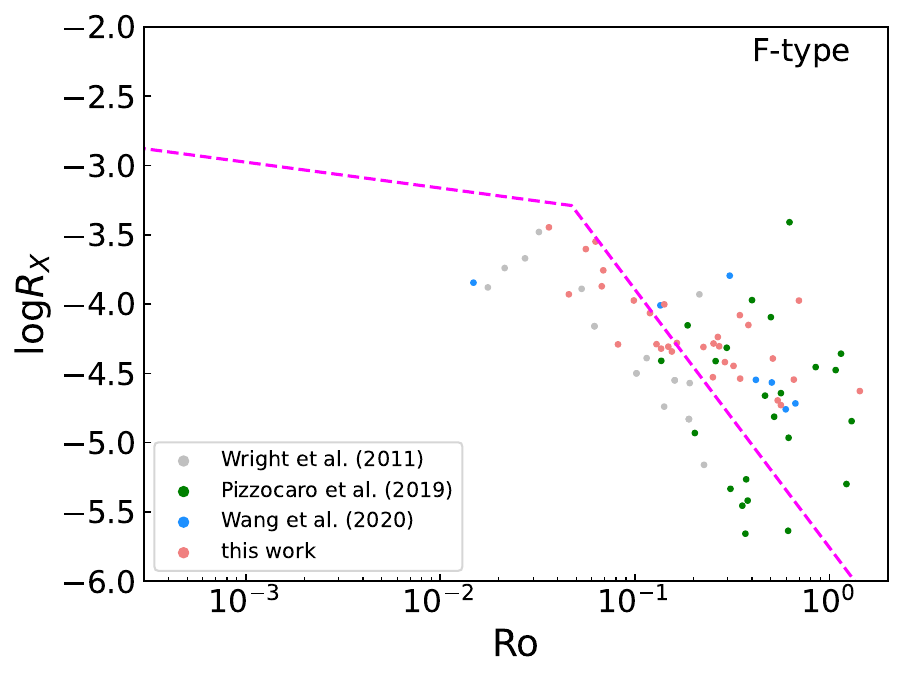}}
%\label{HRD.fig}
\subfigure[]{
\includegraphics[width=0.47\textwidth]{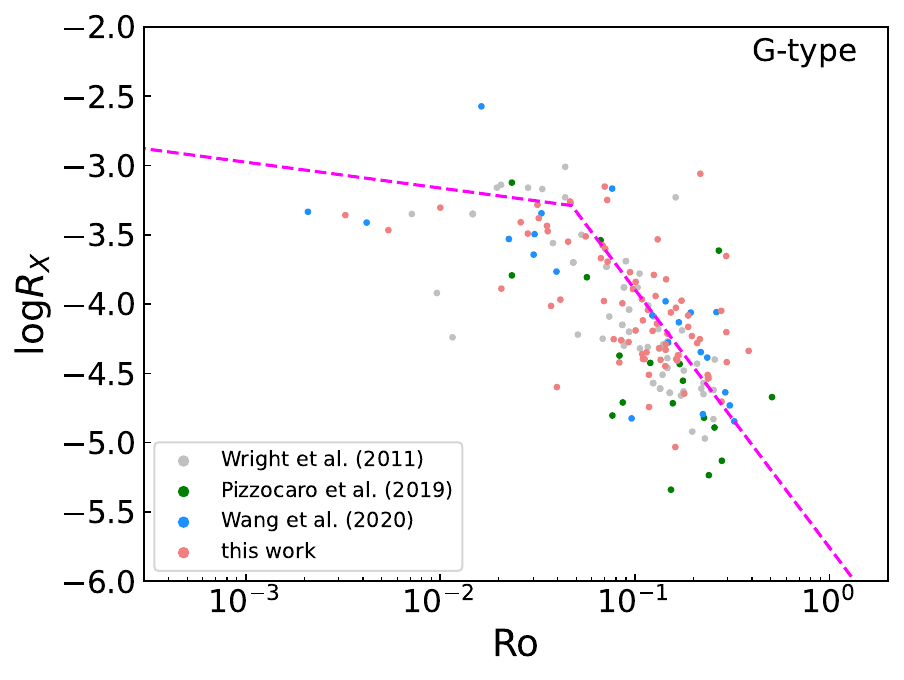}}
\subfigure[]{
\includegraphics[width=0.47\textwidth]{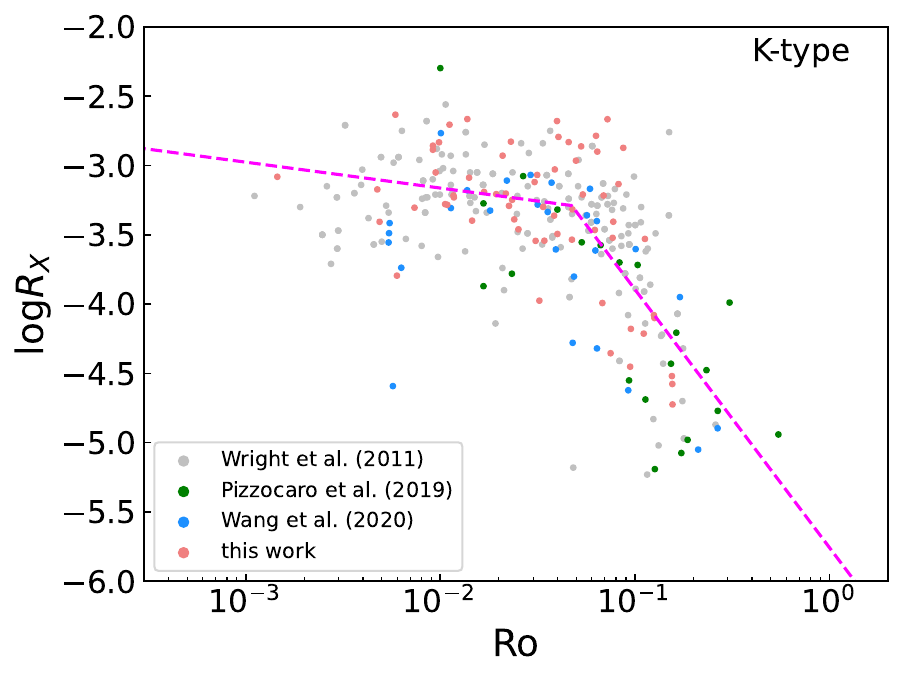}}
\subfigure[]{
\includegraphics[width=0.47\textwidth]{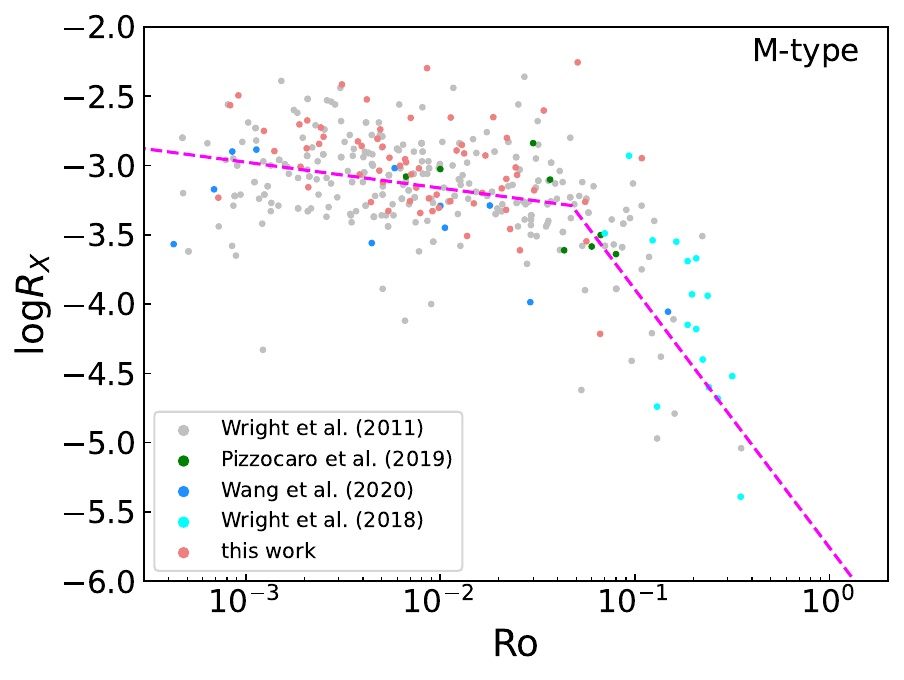}}
%\label{$R_X$.fig}
\caption{%Panel (a), panel (b), panel (c) and panel (d): 
Activity-rotation relations of F-, G-, K-, and M-type single stars, respectively. Stars with $T_{\rm{eff}}$ lower than 4000 are classified as M-type stars. K-type stars are those with $T_{\rm{eff}}$ between 4000K and 5300K. G-type stars have $T_{\rm{eff}}$ ranging from 5300K to 6000K. $T_{\rm{eff}}$ range of F-type stars is set to be from 6000K to 7500K. Magenta dashed line is the best-fit activity-rotation relation shown in Figure \ref{relation}.}
\label{relation_sep}
\end{figure*}

\subsection{Rotation periods and Rossby number}

Rotation is one of the key ingredients in the generation of stellar magnetic field. 
In order to investigate the X-ray activity-rotation relation \citep{2003A&A...397..147P, 2011ApJ...743...48W}, it's necessary to build a sample with accurate rotating period measurements.

Through matching the catalogs mentioned in Section \ref{class.sec}, periods are available for different types of variable stars, including rotation variables, RS CVn variables, and BY Dra variables. The TESS satellite has provided hundreds of thousands of stellar light curves, enabling the expansion of our rotational sample. We cross-matched our sample with the TESS input catalog \citep{2019AJ....158..138S} and downloaded the PDC light curves provided by the TESS science processing operations center (SPOC) \citep{2016SPIE.9913E..3EJ}. TESS light curves used in this work can be found at MAST \citep{https://doi.org/10.17909/t9-nmc8-f686, 2015JATIS...1a4003R}. For each target, the light curves from multiple sectors were combined for period estimation using the Lomb-Scargle (LS) algorithm \citep{1976Ap&SS..39..447L, 1982ApJ...263..835S}.
Finally, the phase-folded light curves were visually checked to ensure reliable period estimation. Furthermore, pulsators were excluded by cross-matching the SIMBAD.

For close binaries, the orbital motion and rotation could be synchronized.
For example, the variation of the light curves of BY Draconis (BY Dra) and RS CVn variables is dominated by the rotation of the late-type giant component, and the rotation periods equals to the orbital periods.
For other binaries such as EA and EB types, the light curves are modulated by orbital motion.

The dimensionless parameter Ro, defined as $P_{\rm{rot}}/\tau_{\rm{c}}$, has been proved to be an appropriate quantity for investigating the stellar activity-rotation relation \citep{1984ApJ...279..763N}.
$\tau_{\rm{c}}$ is the convective turnover time, which was extracted from the Yale-Potsdam Stellar Isochrones \citep[YAPSI;][]{2017ApJ...838..161S}.
For each metallicity grid (i.e., [Fe/H] $=$ +0.3, 0.0, $-$0.5, $-$1.0, $-$1.5), we selected the models within the errors of $T_{\rm{eff}}$ and log$g$, and calculated the median value of $\tau_{\rm{c}}$.
For each star, the final $\tau_c$ value was determined by linear interpolation to the metallicity.
The periods and $\tau_{\rm{c}}$ are listed in Table \ref{tab:table2}.

\subsection{Activity-rotation relation: ``$D^{2}$" model}

The relation between X-ray emission and stellar rotation has long been a hot topic. \cite{1981ApJ...248..279P} found that for early-type stars, $L_X$ is correlated with spectral type (i.e., stellar temperature), while for late-type stars, $L_X$ is strongly proportional to stellar rotation velocity. 
Currently, the activity-rotation relation is described by a piece-wise model, including a saturated region where X-ray activity keeps unchanged and a decay region where X-ray activity is proportional to rotation period or Ro \citep[e.g.][]{2003A&A...397..147P, 2008ApJ...687.1264M, 2011ApJ...743...48W}.

\begin{table*}
%\begin{table*}[h!]
  \centering
  %\begin{center}
    \caption{Derived stellar parameters of each target in our sample.}
    \label{tab:table2}
    \begin{tabular}{cccccccc}
    %\begin{tabular}{p{3cm}<{\centering}p{1.6cm}<{\centering}p{1.25cm}<{\centering}p{1.25cm}<{\centering}p{1.6cm}<{\centering}p{1.25cm}<{\centering}p{1.25cm}<{\centering}p{1.25cm}<{\centering}p{1.25cm}<{\centering}p{1.25cm}<{\centering}p{1.25cm}<{\centering}p{1.25cm}<{\centering}p{1.25cm}<{\centering}}
      \toprule % <-- Toprule here
       Gaia ID & 
       $E(B-V)$ &
       flux &
       $R_X$ &
       $\tau_{c}$ &
       ${P_{\rm{rot}}}$\\
       & (mag) & ($\rm{erg/cm^{2}s}$) &  & (day) & (day)\\
       (1) & (2) & (3) & (4) & (5) & (6)\\
      \hline \\ % <-- Midrule here
158364034415872 & 0.08 & 9.70e-13$\pm$3.80e-13 & 1.10e-01$\pm$4.77e-02 & - & -\\
7798908990819968 & 0.34 & 6.53e-13$\pm$3.22e-13 & 2.24e-04$\pm$1.23e-04 & - & -\\
9124679495764864 & 0.01 & 1.71e-13$\pm$6.15e-14 & 3.68e-04$\pm$1.49e-04 & - & -\\
9770058461712000 & 0.01 & 1.72e-12$\pm$1.74e-13 & 8.62e-04$\pm$1.71e-04 & - & -\\
9964981257351936 & 0.01 & 2.15e-13$\pm$6.10e-14 & 5.21e-05$\pm$1.58e-05 & 14.68 & 1.25\\
17644790803959552 & 0.26 & 5.31e-13$\pm$2.63e-13 & 1.03e-03$\pm$5.20e-04 & 68.33 & 4.7\\
18211043587721088 & 0.01 & 3.50e-12$\pm$3.14e-13 & 8.54e-04$\pm$9.88e-05 & 142.31 & 4.48\\
20896222781456768 & 0.01 & 3.45e-13$\pm$1.06e-13 & 1.18e-04$\pm$4.07e-05 & 18.02 & 0.82\\
24331990460013568 & 0.1 & 6.45e-13$\pm$3.47e-13 & 4.17e-03$\pm$2.42e-03 & - & -\\
27487726270657536 & 0.01 & 7.14e-13$\pm$1.05e-13 & 3.78e-04$\pm$1.19e-04 & - & -\\
29835488537920768 & 0.17 & 1.56e-12$\pm$5.44e-13 & 1.33e-03$\pm$4.73e-04 & - & -\\
... & ... & ... & ... & ... & ... \\
      \hline \\% <-- Bottomrule here
    \end{tabular}\\
  %\end{center}
%\end{table*}
\footnotesize{(1) Gaia ID: Gaia eDR3 source ID. (2) $E(B-V)$: reddening. (3) flux: X-ray flux. (4) $R_X$: normalized X-ray luminosity. (5) $\tau_{c}$: convective turnover time. (6) ${P_{\rm{rot}}}$: rotation period.}
\end{table*}

To investigate X-ray activity-rotation in a large sample, we collected objects with X-ray activity and Ro from previous literature, including \citep{2011ApJ...743...48W, 2018MNRAS.479.2351W, 2019A&A...628A..41P, 2020ApJ...902..114W}. 
The $R_X$ values from \cite{2020ApJ...902..114W} were converted into the energy band of 0.1--2.4 keV
with PIMMS. For studies using empirical $\tau$ \citep[e.g.,][]{2011ApJ...743...48W, 2019A&A...628A..41P, 2018MNRAS.479.2351W}, their Ro values were shifted by Ro/3. 
In our sample, dwarfs with measured rotation periods have log$L_X$ and log$R_X$ ranges of from 28.02 to 30.93 and from -5.03 to -2.25, respectively. The large sample includes 1014 sources.

We applied a piece-wise power law model to fit the activity-rotation relation using the large sample (Figure \ref{relation}) based on the \emph{emcee} python package. It is widely used for Markov Chain Monte Carlo sampling \citep{2013PASP..125..306F}. The model is described as follows,
\begin{equation}
\label{single.eq}
 \log R_X = 
\begin{cases}
 a + b\rm{log}{\rm Ro}, 
 \quad for\ {\rm Ro}\leq{\rm Ro}_{\rm{k}}\\
 a + (b - c)\rm{log}{\rm Ro}_{\rm{k}} + c\rm{log}{\rm Ro},
 \quad for\ {\rm Ro}>{\rm Ro}_{\rm{k}},
 \end{cases}
\end{equation}
where Ro$_{\rm k}$ represents the Ro value at the knee point separating the saturated and unsaturated regimes.
Table \ref{tab:table3} lists the best-fit parameters for single dwarfs, with the slopes being -0.19 in the fast rotating region, -1.86 in the slow rotating region and the break point being -1.32. Posterior distributions of parameters are given in Figure \ref{MCMC}. We named the model consisting of two piece-wise power laws as a ``D$^{2}$" model. This model includes a slow decaying part and a fast decay part, with the knee around Ro $\approx$ 0.047.

\begin{table*}
%\begin{table*}[h!]
  \centering
  %\begin{center}
    \caption{Best-fit parameters of activity-rotation relations.}
    \label{tab:table3}
    \begin{tabular}{ccccc}
    %\begin{tabular}{p{2cm}<{\centering}p{2cm}<{\centering}p{2cm}<{\centering}p{2cm}<{\centering}p{2cm}<{\centering}}
      \toprule % <-- Toprule here
         & 
       a &
       b &
       c &
       $\rm{Ro}_{k}$\\
       
      \hline \\ % <-- Midrule here
      Single stars & $-3.54^{+0.16}_{-0.15}$ & $-0.19^{+0.07}_{-0.07}$ & $-1.86^{+0.21}_{-0.21}$ & $-1.32^{+0.06}_{-0.07}$\\
      Binaries & 
      $-3.75^{+0.34}_{-0.31}$ & $-0.35^{+0.18}_{-0.17}$ & $-2.07^{+0.25}_{-0.29}$ & $-1.32^{+0.1}_{-0.1}$\\
      \hline \\
% <-- Bottomrule here
    \end{tabular}\\
  %\end{center}
%\end{table*}
%\footnotesize{(1) Gaia ID: Gaia eDR3 source ID. (2) E(B-V): reddening (3) flux: X-ray flux. (4) $R_X$: normalized X-ray luminosity. (5) $\tau_{c}$: convective turnover time. (6) $\rm{P_{rot}}$: rotation period.}
\end{table*}

In contrast to \cite{2011ApJ...743...48W}, our study reveals that in the classical saturation region, $R_X$ continues to increase as stars rotate faster (i.e., in the first decay region of the ``D$^{2}$" model),
a trend which has also been noticed in previous studies \citep{2014ApJ...794..144R, 2021A&A...649A..96J}.
Recently, based on the newly published \emph{eROSITA} data, \cite{2024arXiv240117282F} used a polynomial model to describe the activity-rotation relation, which also shows an increase of $R_X$ in relation to the decreasing Ro (Figure \ref{relation}).
In order to examine the difference between the relations given by \cite{2024arXiv240117282F} and other studies, we re-calculated the $R_X$ values of \cite{2024arXiv240117282F} using their catalog.
Figure \ref{cross_various.fig} shows a good agreement of $R_X$ between these studies, suggesting the different relation presented by \cite{2024arXiv240117282F} may be due to their estimation method of $\tau$.

In the saturated regime, theoretical work has proposed that as a star rotates faster, the mixing length of a turbulent eddy will gradually grows until it reaches the depth of convective zone, which could lead to the saturation of magnetic energy \citep{2022ApJ...926...40W}. 
However, fast rotating stars tend to be young stars, which could develop deeper convective zone. This would allow the continual growth of mixing length and thus the magnetic energy (towards younger ages). 
\cite{2014ApJ...794..144R} suggested that the trend may be due to the remaining dependence of dynamo on rotation in this region or the scatter of X-ray luminosity of different stellar masses. Figure \ref{relation_sep} shows that the saturation region is mainly occupied by K and M dwarfs. For both K and M dwarfs, especially M dwarfs, the decay trend can be clearly seen.
This suggests the trend is most possibly due to a remaining dependence of magnetic dynamo on rotation, rather than a scatter of X-ray luminosities of different types of stars.

\subsection{Fine structures in the activity-rotation relation: ``SD$^2$F" model}

\begin{figure}
\includegraphics[width=0.47\textwidth]{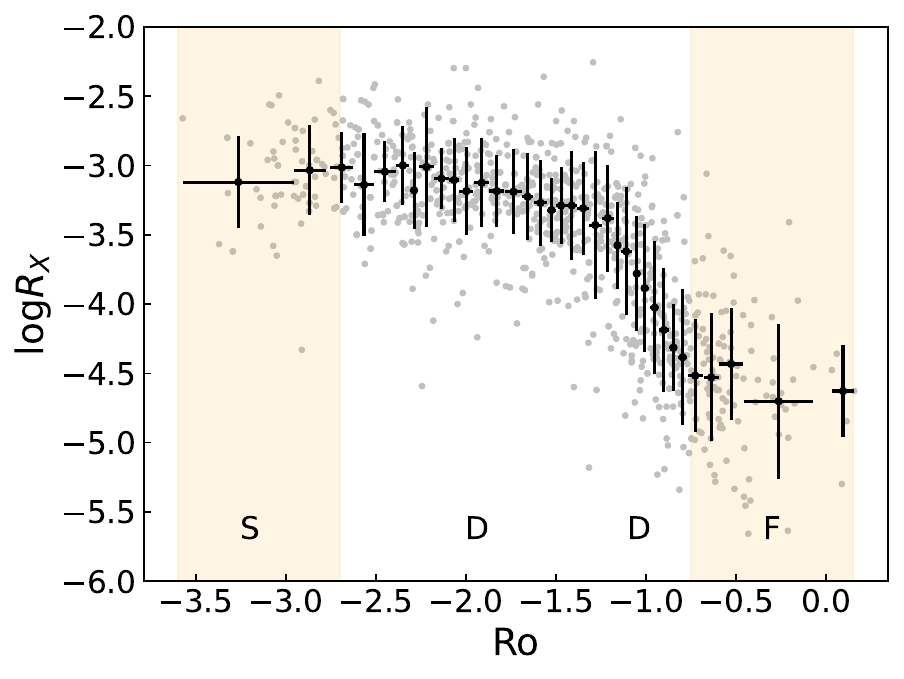}
\caption{Fine structures of activity-rotation relation of single stars. ``S" represents the super-saturation region. The first ``D" represents the weak decay region while the second one corresponds to the fast decay region. ``F" stands for the region with an almost flat slope.}
\label{fine_structure}
\end{figure}

Although the ``D$^{2}$" model could generally describe the overall trends in the activity-rotation relation, there are still some fine structures in the low Ro region and high Ro region. 
We divided the sample into different bins of Ro, each containing 30 targets. We calculated the median value of $R_X$ and the standard deviation for each bin.
Figure \ref{fine_structure} shows a decline of $R_X$ when logRo is smaller than $-$2.7 and a significantly reduced slope when logRo is larger than $-$0.7.
Therefore, the relation can be described by a ``SD$^2$F" model, including one super-saturated region, two decay regions with different slopes and one flat region with almost constant $R$x in the slow rotating regime.

In our study, the super-saturation is mainly caused by M stars.
The super-saturation phenomenon has been observed in many previous studies on stellar X-ray activity \citep{1996A&A...305..785R, 1997ApJ...479..776S, 2011ApJ...743...48W} and still remains a topic of debate.
Possible explanations include the decreasing of filling factor of active regions \citep{2001A&A...370..157S} or the centrifugal stripping of stellar corona, which could result in the decreasing of X-ray emitting volume \citep{2000MNRAS.318.1217J}. 
Recently, our recent work has found some evidences of super-saturation with the $R^{\prime}_{\rm{HK}}$ (Ding et al. in prep); \citet{2022MNRAS.514.4932C} found a decline in the saturation value of $S_{\rm ph}$ index for BY Dra variables with short periods.
The super-saturation observed in X-ray, chromospheric, and photospheric activity proxies would favor the scenario of the decrease in the filling factor of active regions as the mechanism for super-saturation.

The flat region (with high logRo) consists of F stars. The significantly reduced slope is caused by the scatter in $R_X$ of F stars. \cite{2019A&A...628A..41P} suggested that among F stars, there may be unresolved active binaries contributing to additional (higher) X-ray activity.
However, no similar scatter is observed for other types of stars. 
Meanwhile, \cite{2023ApJS..264...12H} also noted a scatter of $R_{\rm H{\alpha}}^{'}$ for F and G stars in the high logRo regime, which can be attributed to the intrinsic variability of stellar activities. 
All the F dwarfs in this study and most F stars in \cite{2019A&A...628A..41P} have $T_{\rm{eff}}$ lower than 6500K, meaning that they are late-type F stars with thin convective zones and possibly rotation-related magnetic dynamos.
It's worthy to investigate whether the scatter is caused by stars with a different dynamo.

\subsection{Activity-rotation relation of binaries}

\begin{figure*}[t]
%\centering
\includegraphics[width=0.5\textwidth]{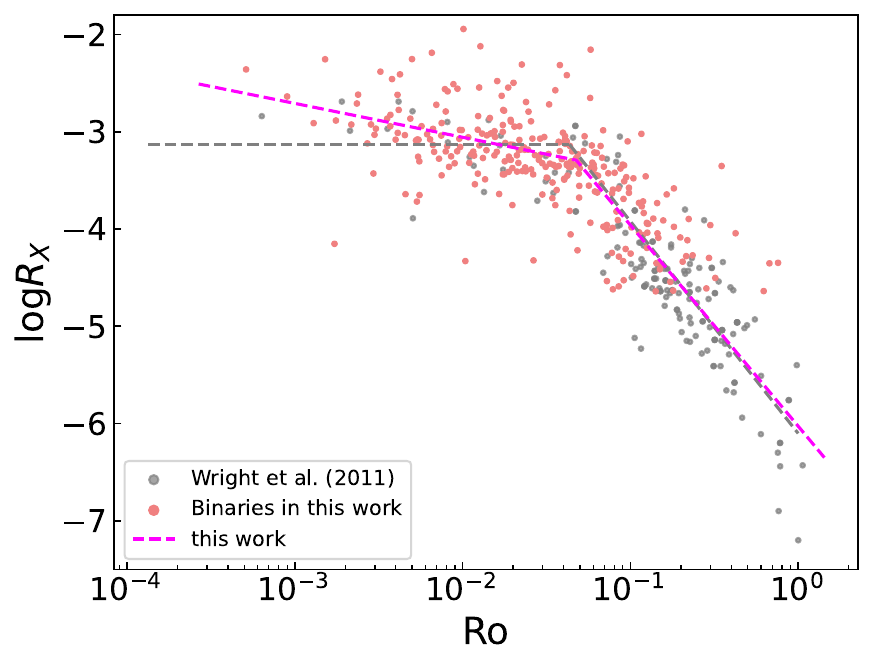}
\includegraphics[width=0.5\textwidth]{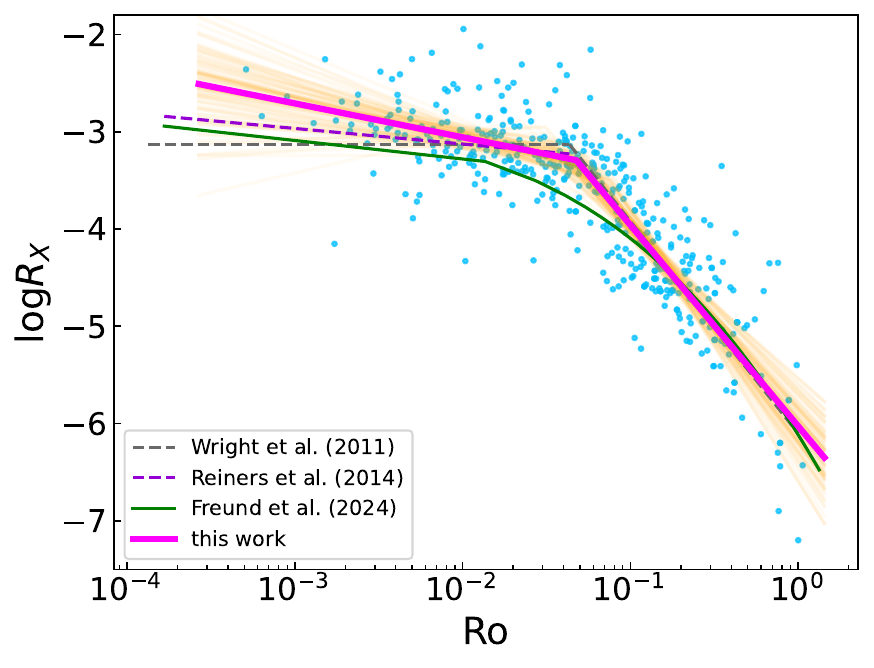}
%\label{$R_X$.fig}
\caption{Panel (a): Activity-rotation relation of binaries. The grey and lightcoral dots are from \cite{2011ApJ...743...48W} and this work, respectively. Ro from \cite{2011ApJ...743...48W} are shifted by a factor of Ro$/$3. The magenta line is our best-fit result. Panel (b): Comparisons of activity-rotation relations from different studies. The shaded region shows 100 models extracted randomly from the posterior probability distribution.}
\label{relation_var}
\end{figure*}

\begin{figure*}
%\centering
\subfigure[]{
\includegraphics[width=0.5\textwidth]{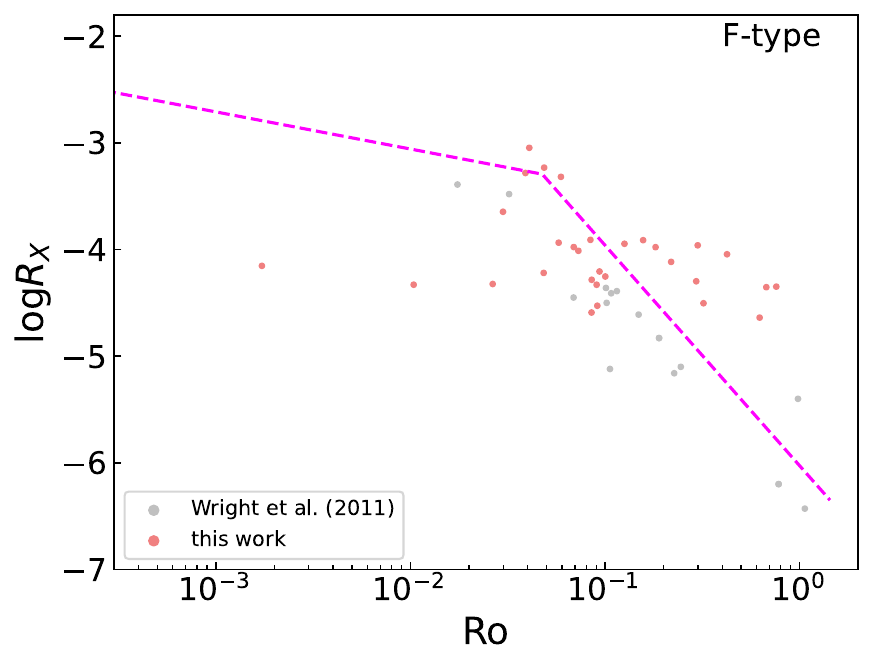}}
\subfigure[]{
\includegraphics[width=0.5\textwidth]{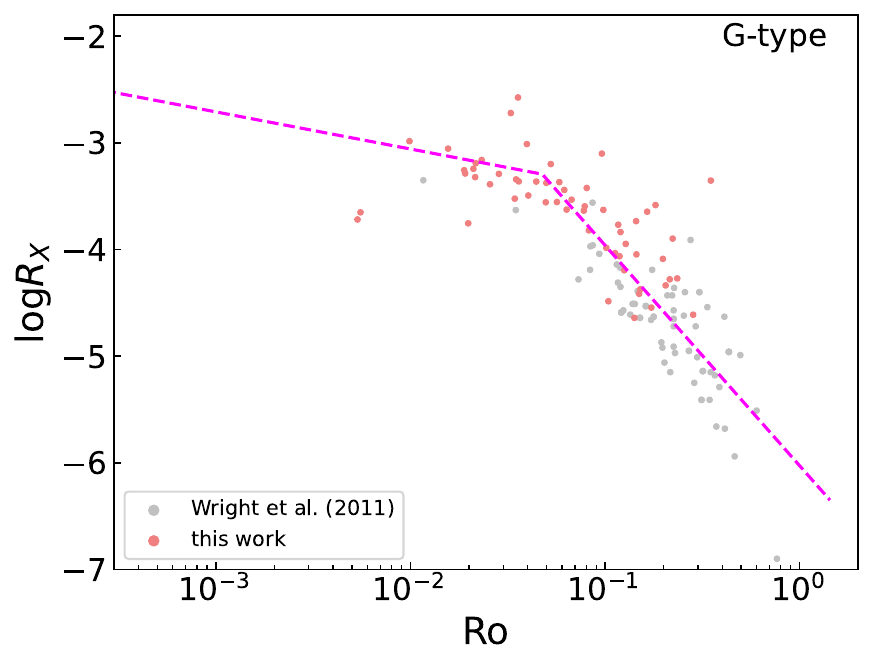}}
\subfigure[]{
\includegraphics[width=0.5\textwidth]{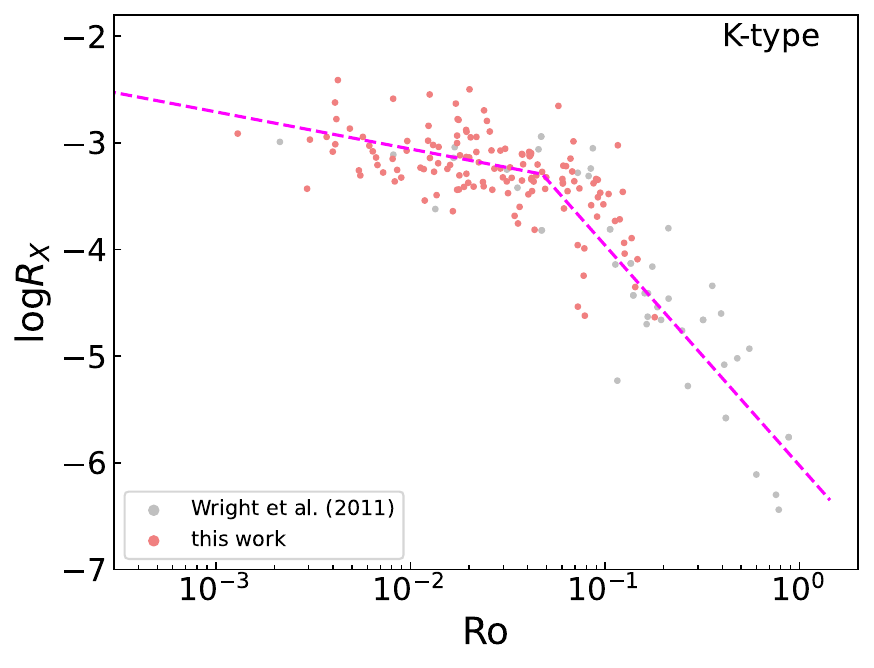}}
\subfigure[]{
\includegraphics[width=0.5\textwidth]{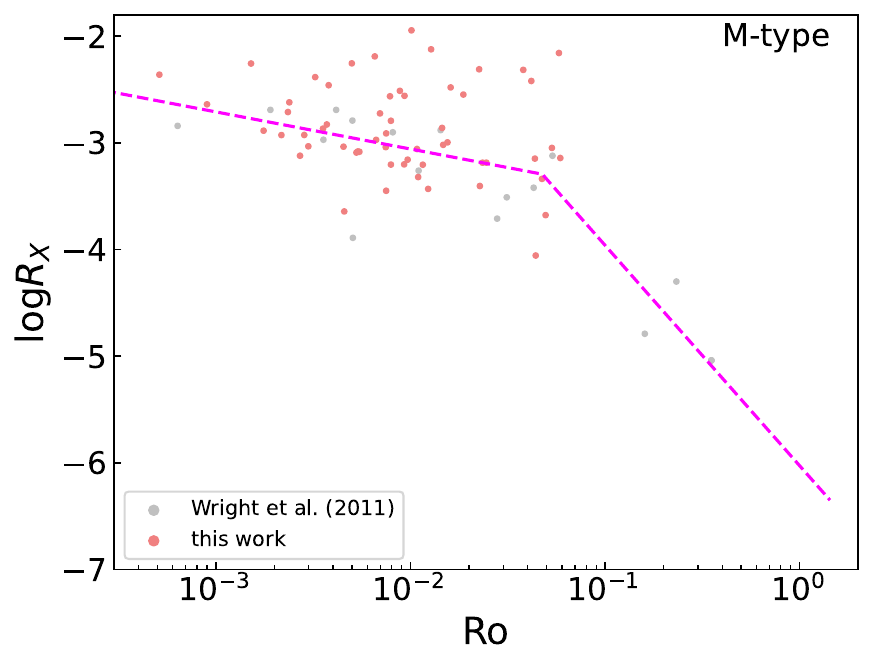}}
\label{$R_X$.fig}
\caption{Activity-rotation of F-, G-, K-, and M-type binary stars, respectively. Binaries are categorized using the same criteria given in the caption of Figure \ref{relation_sep}. Magenta dashed line is the best-fit activity-rotation relation shown in Figure \ref{relation_var}.}
\label{relation_sep_binary}
\end{figure*}

Binaries show similar activity-rotation relation to single stars(Figure \ref{relation_var} and Figure \ref{relation_sep_binary}).
In our sample, binaries with measured rotation periods have log$L_X$ and log$R_X$ ranges of from 28.19 to 30.99 and from $-$4.64 to $-$1.94, respectively.
The fitting results with a two-piece power law are given in Table \ref{tab:table3}, with the slopes being $-$0.35 and $-$2.07 in the fast rotating region and slow rotating region, respectively. The knee point is logRo $=-$1.32. Meanwhile, some binaries could exhibit higher activity levels compared to single stars due to tidal interaction.
There are 15 binaries with log$R_X$ larger than $-$2.5. 
Among these sources, one was classified as RS CVn variables and six were classified as BY Dra variables, both of which are close binaries.
The activity level affected by tidal interaction may explain the large scatter shown in the classical saturation region of the binary activity-rotation relation.

\section{The Ultraluminous sample}
\label{ultraluminous.sec}

\begin{figure*}
\centering
\subfigure[]{
\includegraphics[width=0.48\textwidth]{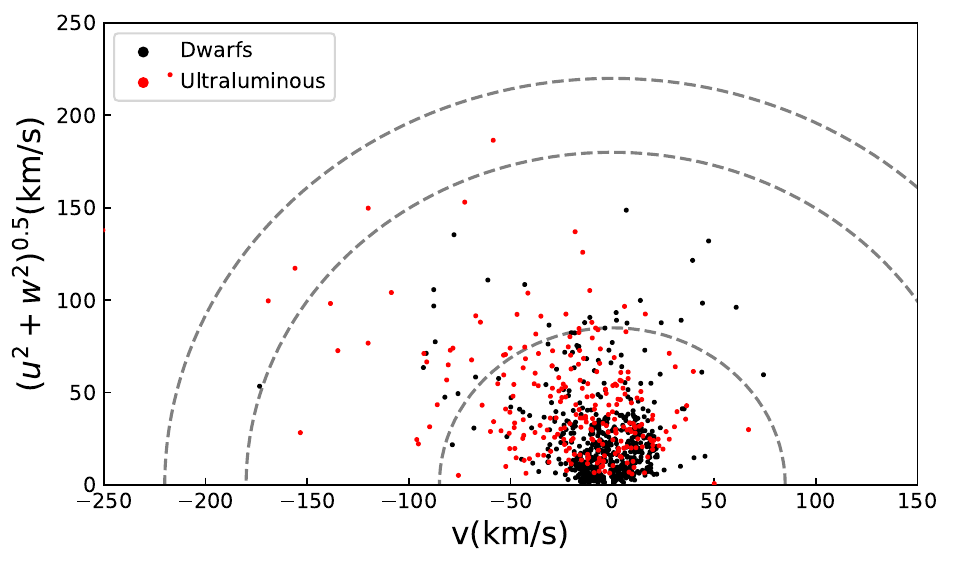}}
\subfigure[]{
\includegraphics[width=0.48\textwidth]{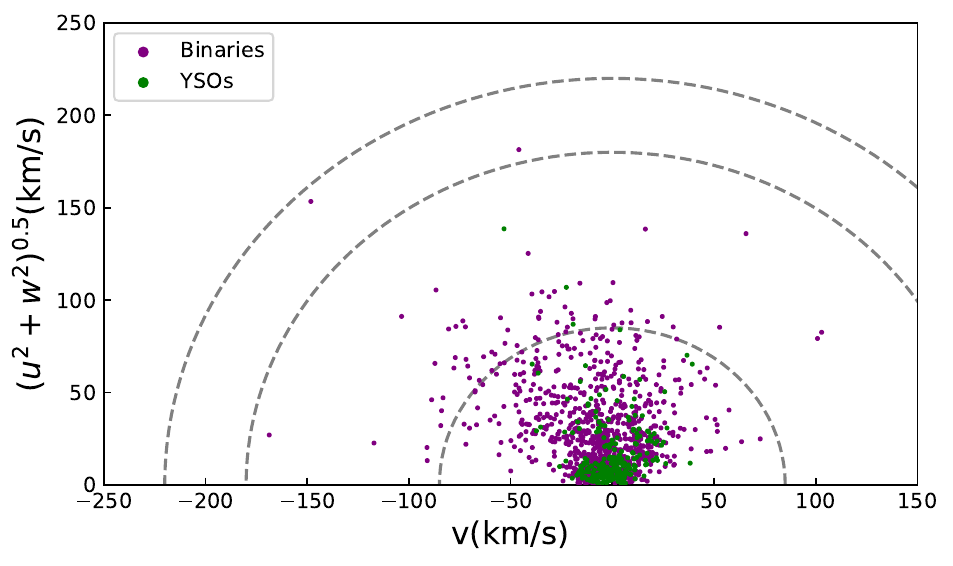}}
\subfigure[]{
\includegraphics[width=0.49\textwidth]{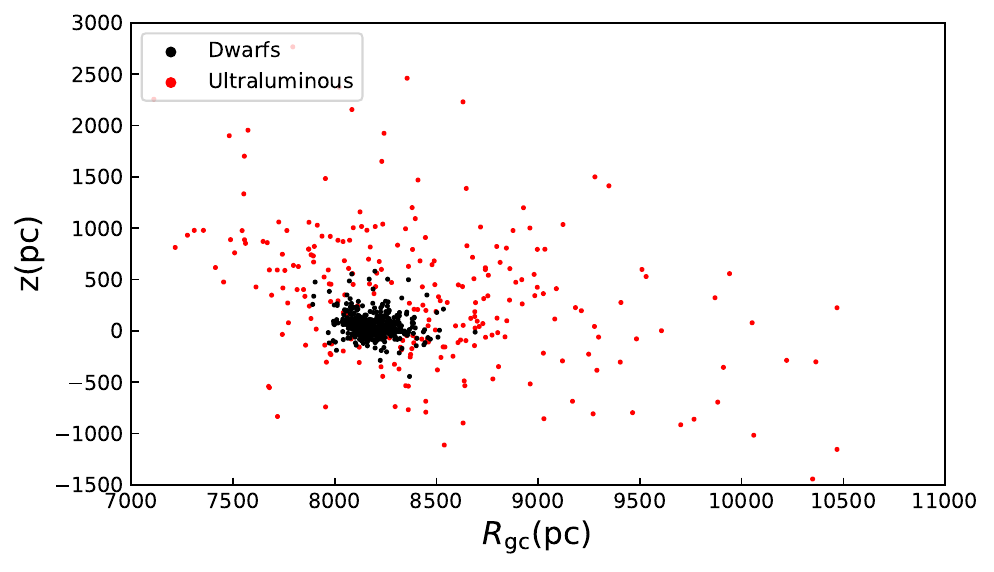}}
\subfigure[]{
\includegraphics[width=0.49\textwidth]{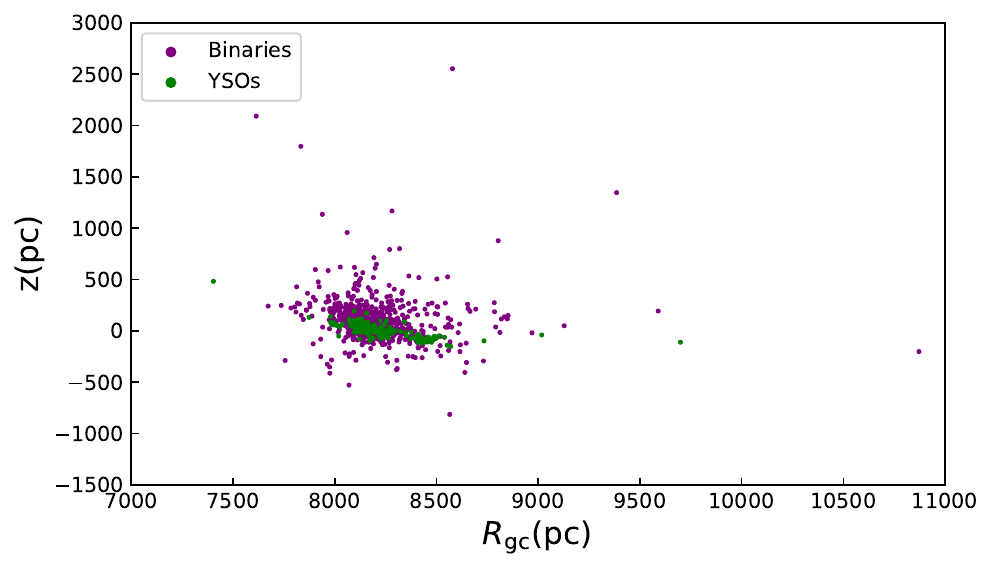}}
\caption{Panel (a): Toomre diagram of dwarfs and Ultraluminous sources. Panel (b): Toomre diagram of binaries and YSOs. (c) Galactic distribution of dwarfs and Ultraluminous sources. $z$ is the vertical distance to the Galactic plane. $R_{\rm{gc}}$ is the radial distance to the Galactic center. (d) Galactic distribution of binaries and YSOs.}
\label{galactic.fig}
\end{figure*}

There is an Ultraluminous population of main-sequence stars with X-ray luminosities ranging from $10^{31}$ to $10^{33}$ erg/s, whose nature warrants investigation.

Although the luminosities of this population are close to the brightest YSOs and binaries, their $R_X$ distribution spans from $-$3 to $-$1, significantly exceeding the distribution of YSOs and binaries. Furthermore, the spatial distribution of this population is much more scattered compared with YSOs and binaries (Figure \ref{galactic.fig}), indicating they are different types of sources.
Therefore, YSOs and binaries may have a minor contribution to the Ultraluminous population.

As discussed in Section \ref{target.sec} and \ref{class.sec}, in our sample, approximately $10\%$ of the optical counterparts to the X-ray sources may be incorrectly identified, due to the poor position accuracy of $ROSAT$.
Previous studies have shown that galaxies and AGNs typically exhibit much higher X-ray to optical flux ratios than stars \citep[e.g.][]{1991ApJS...76..813S, 2003AJ....126..575H}.
Recently, through comparing the optical counterparts identified by different methods, \cite{2024arXiv240117282F} found a group of stars with distinctly high X-ray to bolometric flux ratios and argued that they are extragalactic sources. Furthermore, it's reasonable to expect that galaxies and AGNs would have different spatial resolutions compared to stars, as shown in Figure \ref{galactic.fig}.
Besides, the portion of the Ultraluminous sources in the sample (i.e., $\sim 13\% = 258/2042$) roughly matches the false match rate. 
We further cross-matched the dwarf sample with the GLADE+ catalog \citep{2022MNRAS.514.1403D} using match radii of 10", 20", and 30", corresponding to 12, 68, and 146 targets, respectively. 
Therefore, we suspected that the Ultraluminous population could mainly consist of extragalactic sources.

\begin{figure*}[t]
%\centering
\subfigure[]{
\includegraphics[width=0.47\textwidth]{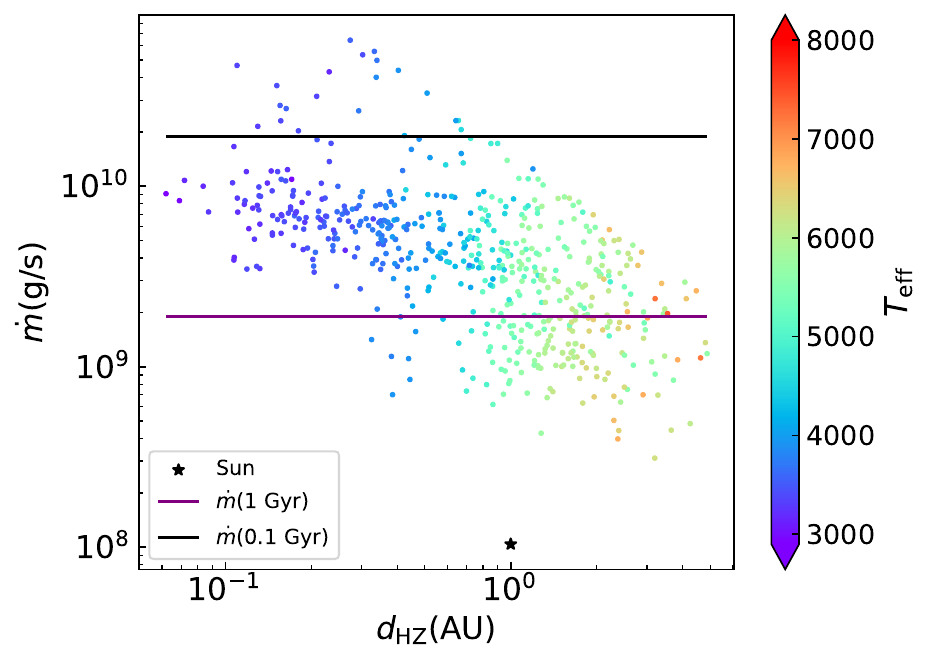}}
%\label{HRD.fig}
\subfigure[]{
\includegraphics[width=0.47\textwidth]{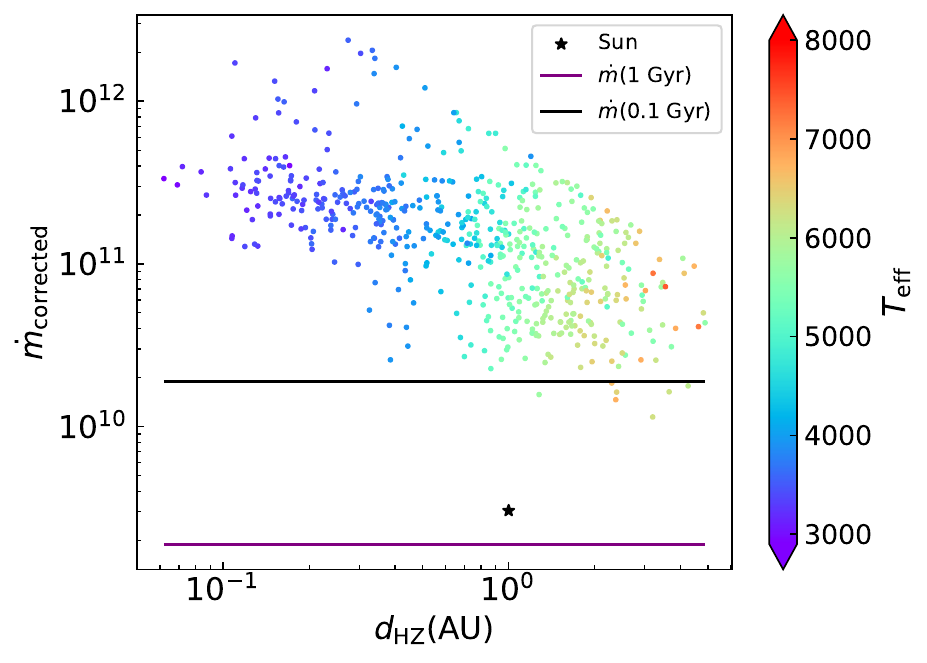}}
%\label{$R_X$.fig}
\caption{Panel (a): Mass loss rate of the planetary primordial atmosphere induced by X-ray emission. The mass loss rate within 0.1 and 1 Gyr are plotted with black and purple horizontal lines, respectively. 
Stars above the lines suggest they can remove the planetary primordial envelope within 0.1 or 1 Gyr. Different colors represent different effective temperatures. The Sun is donated by black star. Panel (b): Mass loss rate of planetary atmosphere induced by stellar X-ray emission which has been corrected to younger age.}
\label{habitable}
\end{figure*}

\section{X-ray emissions and habitable zone}
\subsection{mass loss rate induced by current X-ray emission}
\label{habit.sec}

The impact of stellar activity on stellar habitability is a hot topic. Generally, the habitable zone of a star is defined as the region in which liquid water can retain on a planet \citep[e.g.][]{1993Icar..101..108K, 2013ApJ...765..131K}. However, the presence of liquid water is just one of the prerequisites for the generation of life. High energy emissions like UV and X-ray radiation could strongly affect the planetary habitability. UV emissions will not only influence the production of $\rm{O_{3}}$ in the planetary atmosphere \citep{2003AsBio...3..689S} but also trigger the formation of organic molecules \citep{2009Natur.459..239P}. 
%Naturally, the concept UV habitable zone has come into being, in which a planet will receive moderate UV radiation \citep{2007Icar..192..582B, 2023MNRAS.522.1411S}.
The UV habitable zone has been defined as the region in which a planet receives moderate UV radiation \citep{2007Icar..192..582B, 2023MNRAS.522.1411S}.

Meanwhile, to establish a suitable temperature and pressure on a planet, the removal of its primordial H/He atmosphere is necessary, and a secondary atmosphere needs to be generated through processes such as volcanic activities to maintain liquid water \citep{2020PNAS..11718264K, 2021AJ....161..213S}. The high energy radiation of host stars plays a key role in the mass loss of the primordial atmosphere due to photoevaporation. 

In this work, following the approaches of \cite{2009ApJ...693...23M} and \cite{2020MNRAS.499...77K}, we investigated the impact of X-ray emissions on the habitability of a planetary. 
First, for each single star in our sample, we calculated the inner boundary($d_{\rm{in}}$) and outer boundary($d_{\rm{out}}$) of the continuous habitable zone using the formular given by \cite{2014ApJ...787L..29K}:
\begin{equation}
d = (\frac{L/L_{\odot}}{S_{\rm{eff}}})^{0.5} \rm{AU},
\end{equation}
where $S_{\rm{eff}}$ is calculated according to the coefficients from \cite{2014ApJ...787L..29K}. 
Second, assuming there is a planet with one Earth mass ($M_{\rm{p}} = 1M_{\otimes}$) and one Earth radius ($R_{\rm{P}} = 1R_{\otimes}$) in the middle region of the continuous habitable zone ($\frac{d_{\rm{out}} + d_{\rm{in}}}{2}$), the mass loss rate $\dot{m}$ induced by X-ray emissions was estimated using the formulae from \cite{2020MNRAS.499...77K}:
\begin{equation}
\dot{m} = \frac{\eta\pi F_{\rm{EUV}}R_{\rm{p}}R_{\rm{XUV}}^{2}}{GM_{\rm{p}}K},
\end{equation}
where $F_{\rm{EUV}}$ is the EUV flux at the distance of $\frac{d_{\rm{out}} + d_{\rm{in}}}{2}$. $\eta$ is the parameter named heat efficiency and was set to be 0.15. 
The $R_{\rm{XUV}}$ is the effective absorption radius of XUV photons, which was defined by \cite{2016ApJ...831..180C}:
\begin{equation}
R_{\rm{XUV}} = R_{\rm{p}} + H\rm{ln}\frac{\emph{P}_{\rm{photo}}}{\emph{P}_{\rm{XUV}}}.
\end{equation}
Here \emph{H} is the atmospheric scale height and $P_{\rm{photo}}$ is the pressure, whose values were adopted from \cite{2009ApJ...693...23M}. $P_{\rm{XUV}}$ is the XUV absorption level.
The factor $K$ takes into account that the atmospheres only escape when it approaches the Roche lobe boundary， which was defined by \cite{2007A&A...472..329E} as:
\begin{equation}
K = 1 - \frac{3}{2\xi} + \frac{1}{2\xi^{3}}
\end{equation}
and
\begin{equation}
\xi = (\frac{M_{\rm{p}}}{3M_{\rm{s}}})^{1/3}\frac{a}{R_{\rm{p}}}.
\end{equation}
$M_{\rm{s}}$ is mass of host star and $a$ is the semi-axis of planet, which was set to be $\frac{d_{\rm{out}} + d_{\rm{in}}}{2}$. The EUV flux was converted from our X-ray flux using the relation from \cite{2015Icar..250..357C}: 
\begin{equation}
\rm{log}\emph{F}_{\rm{EUV}} = 2.63 + 0.58\rm{log} \emph{F}_{\rm{x}}
\end{equation}
For detailed descriptions of above parameters we refer to \cite{2009ApJ...693...23M}, \cite{2016ApJ...831..180C}, and \cite{2020MNRAS.499...77K}.

If a planet keeps its primordial H/He atmosphere for a significant time, even though it resides in habitable zone, it won't be habitable \citep{2016MNRAS.459.4088O}. A rocky Earth-like planet could accrete H/He envelope with a mass fraction of 0.1$\%$ to 1$\%$, and such a scenario has been suggested by both theoretical and observational studies \citep[e.g.][]{2015ApJ...806..183W, 2011ApJ...738...59R}. Assuming there is a planet with a radius of $1R_{\otimes}$ and a primordial envelope fraction of $1\%$ in the habitable zone of each star in our sample, we examined whether the X-ray-induced mass loss rate of planetary atmosphere is sufficient to significantly remove primordial envelope over 0.1 or 1 Giga years. 

The mass loss rate to the primordial envelope within 0.1 Gyr and 1 Gyr are plotted by black and purple horizontal lines in Figure \ref{habitable}, respectively. 
Stars above the lines suggest their mass loss rates are high enough to remove the primordial envelope.
Distances of the continuous habitable zones to M dwarfs are small, leading to much more significant atmospheric stripping due to X-ray emissions. 
Figure \ref{habitable} suggests the primordial envelope of most of the M dwarfs could be removed within 1 Gyr, which is consistent with the results of \cite{2016MNRAS.459.4088O}.

\subsection{mass loss rate corrected to young ages of stars}

It's strange that our Sun lies below these lines, indicating an insufficient mass loss rate of the primordial envelope.
One scenario is that the sun may have exhibited a much higher X-ray luminosity during the early formation stage of solar system \citep{1997ApJ...483..947G}.
Therefore, it would be worthwhile to correct the current X-ray luminosities of these stars to the values for their youth.

Previous studies have shown that the levels of stellar chromospheric and coronal activities would decrease as stars age \citep[e.g.][]{2008ApJ...687.1264M}. Recently, \cite{2023ApJ...951...44R} suggested the X-ray fluxes of K- and M-type stars can decrease by 100 to 1000 times when stars age from 10 Myr to 100 Myr.
Therefore, the planetary atmosphere may suffer from a much stronger mass loss rate during the early stage of its host star.

Here we re-calculated the mass loss rate by using an X-ray luminosity corresponding to the stars' youth for our sample stars.
The X-ray luminosity during the youth was assumed to be 500 times higher than the current value. 
Figure \ref{habitable} shows that with the re-calculated mass loss rate, most of the dwarfs could loss the primordial H/He envelope within 0.1 Giga year. Meanwhile, the X-ray emission from young Sun can strip away the Earth's primordial envelope with in 1 Giga year, thus enabling the development of life.

\section{Summary}

In this work, utilizing the homogeneous \emph{ROSAT} 2RXS all-sky X-ray catalog, together with LAMOST DR10, APOGEE DR16, Gaia eDR3 and various photometric sky surveys, we performed a comprehensive estimation of X-ray activities of late-type stars.
Our sample includes a total of 2042 targets, among which 521 targets exhibit periodic modulation in light curves. The X-ray activity was quantified by the normalized X-ray flux $R_X$ = $f_X/f_{\rm bol}$. A detailed study of stellar activity-rotations was carried out for both single stars and binaries.

The overall activity-rotation relation can be roughly divided into two parts: a weak decay region and a rapid decay region, which can be referred to as the ``D$^{2}$" model. In the fast-rotating regime, known as the saturated regime, our results show that there is still a weak dependence of $R_X$ on Ro, suggesting a continued dependence of magnetic dynamo on stellar rotation. 
This contrasts with the conventional picture, which has shown a constant value of the activity index in this region.

However, detailed studies revealed some fine structures in the activity-rotation relation. In the extremely fast rotating regime, mainly composed of M stars, $R_X$ was observed to decrease when stars rotate faster, a phenomenon named super-saturation, suggesting the decrease of the filling factors of active regions.
On the other hand, in the extremely slow rotating region, a flat slope of $R_X$ towards Ro was observed, mainly attributed to the scattering of F stars. This phenomenon may be caused by intrinsic variability of stellar activities (during one stellar cycle) or different magnetic dynamos. We named the detailed activity-rotation relation, which can be divided into four parts, the ``SD$^2$F" model.

Meanwhile, the activity-rotation relation of binaries is similar to that of single stars, suggesting the X-ray activities of binaries are dominated by the more active components.
A slightly larger slope was observed in the fast rotating region for binaries, which can be attributed to the enhanced activity due to tidal interaction as binaries approach close separation.
However, due to the limited sample size, no fine structure was found in the relation for binaries.

A group of dwarfs with very high $L_X$ and $R_X$ values were identified as the Ultraluminous sources. The size of the Ultraluminous sample roughly matches the false match rate of our sample. 
These sources exhibit a different spatial distribution compared to other stars, such as dwarfs, giants, and YSOs. 
Therefore, we suspected the Ultraluminous sample may mainly consist of extragalactic sources, such as AGN and galaxies.

Finally, we investigated the impacts of X-ray emissions to planetary atmosphere. By calculating the mass loss rate of planetary primordial envelope in the continuous habitable zone, we found that most of the dwarfs can remove the primordial H/He envelope, which accounts for approximately 1\% of the planetary mass, within one Giga year, especially for M-dwarfs whose habitable zones are close to host stars.

\section*{acknowledgements}

The Guoshoujing Telescope (the Large Sky Area Multi-Object Fiber Spectroscopic Telescope LAMOST) is a National Major Scientific Project built by the Chinese Academy of Sciences. Funding for the project has been provided by the National Development and Reform Commission. LAMOST is operated and managed by the National Astronomical Observatories, Chinese Academy of Sciences.
Some of the data presented in this paper were obtained from the Mikulski Archive for Space Telescopes (MAST).
This work presents results from the European Space Agency (ESA) space mission {\it Gaia}. {\it Gaia} data are being processed by the {\it Gaia} Data Processing and Analysis Consortium (DPAC). Funding for the DPAC is provided by national institutions, in particular the institutions participating in the {\it Gaia} MultiLateral Agreement (MLA). The {\it Gaia} mission website is https://www.cosmos.esa.int/gaia. The {\it Gaia} archive website is https://archives.esac.esa.int/gaia. We acknowledge use of the VizieR catalog access tool, operated at CDS, Strasbourg, France, the \emph{corner} package from \cite{corner} and of Astropy, a community-developed core Python package for Astronomy (Astropy Collaboration, 2013).
This work was supported by National Key Research and Development Program of China (NKRDPC) under grant Nos. 2019YFA0405000 and 2019YFA0405504, Science Research Grants from the China Manned Space Project with No. CMS-CSST-2021-A08, Strategic Priority Program of the Chinese Academy of Sciences undergrant No. XDB4100000, and National  Natural Science Foundation of China (NSFC) under grant Nos. 11988101/11933004/11833002/12090042/12273057. S.W. acknowledges support from the Youth Innovation Promotion Association of the CAS (IDs 2019057).

\bibliographystyle{yahapj}
\bibliography{main}

\begin{thebibliography}{}
\expandafter\ifx\csname natexlab\endcsname\relax\def\natexlab#1{#1}\fi
\providecommand{\url}[1]{\href{#1}{#1}}

\bibitem[{{Alfv{\'e}n}(1947)}]{1947MNRAS.107..211A}
{Alfv{\'e}n}, H. 1947, \mnras, 107, 211

\bibitem[{{Bailer-Jones} {et~al.}(2021){Bailer-Jones}, {Rybizki}, {Fouesneau},
  {Demleitner}, \& {Andrae}}]{2021AJ....161..147B}
{Bailer-Jones}, C.~A.~L., {Rybizki}, J., {Fouesneau}, M., {Demleitner}, M., \&
  {Andrae}, R. 2021, \aj, 161, 147

\bibitem[{{Boller} {et~al.}(2016){Boller}, {Freyberg}, {Tr{\"u}mper}, {Haberl},
  {Voges}, \& {Nandra}}]{2016A&A...588A.103B}
{Boller}, T., {Freyberg}, M.~J., {Tr{\"u}mper}, J., {et~al.} 2016, \aap, 588,
  A103

\bibitem[{{Buccino} {et~al.}(2007){Buccino}, {Lemarchand}, \&
  {Mauas}}]{2007Icar..192..582B}
{Buccino}, A.~P., {Lemarchand}, G.~A., \& {Mauas}, P. J.~D. 2007, \icarus, 192,
  582

\bibitem[{{Chadney} {et~al.}(2015){Chadney}, {Galand}, {Unruh}, {Koskinen}, \&
  {Sanz-Forcada}}]{2015Icar..250..357C}
{Chadney}, J.~M., {Galand}, M., {Unruh}, Y.~C., {Koskinen}, T.~T., \&
  {Sanz-Forcada}, J. 2015, \icarus, 250, 357

\bibitem[{{Chahal} {et~al.}(2022){Chahal}, {de Grijs}, {Kamath}, \&
  {Chen}}]{2022MNRAS.514.4932C}
{Chahal}, D., {de Grijs}, R., {Kamath}, D., \& {Chen}, X. 2022, \mnras, 514,
  4932

\bibitem[{{Chen} \& {Rogers}(2016)}]{2016ApJ...831..180C}
{Chen}, H., \& {Rogers}, L.~A. 2016, \apj, 831, 180

\bibitem[{{Chen} {et~al.}(2018){Chen}, {Wang}, {Deng}, {de Grijs}, \&
  {Yang}}]{2018ApJS..237...28C}
{Chen}, X., {Wang}, S., {Deng}, L., {de Grijs}, R., \& {Yang}, M. 2018, \apjs,
  237, 28

\bibitem[{{Chen} {et~al.}(2020){Chen}, {Wang}, {Deng}, {de Grijs}, {Yang}, \&
  {Tian}}]{2020ApJS..249...18C}
{Chen}, X., {Wang}, S., {Deng}, L., {et~al.} 2020, \apjs, 249, 18

\bibitem[{{Christy} {et~al.}(2022){Christy}, {Jayasinghe}, {Stanek},
  {Kochanek}, {Way}, {Prieto}, {Shappee}, {Holoien}, {Thompson}, \&
  {Schneider}}]{2022PASP..134b4201C}
{Christy}, C.~T., {Jayasinghe}, T., {Stanek}, K.~Z., {et~al.} 2022, \pasp, 134,
  024201

\bibitem[{{Cui} {et~al.}(2012){Cui}, {Zhao}, {Chu}, {Li}, {Li}, {Zhang}, {Su},
  {Yao}, {Wang}, {Xing}, {Li}, {Zhu}, {Wang}, {Gu}, {Luo}, {Xu}, {Zhang},
  {Liu}, {Zhang}, {Yang}, {Cao}, {Chen}, {Chen}, {Chen}, {Chen}, {Chu}, {Feng},
  {Gong}, {Hou}, {Hu}, {Hu}, {Hu}, {Jia}, {Jiang}, {Jiang}, {Jiang}, {Jin},
  {Li}, {Li}, {Li}, {Liu}, {Liu}, {Lu}, {Mao}, {Men}, {Qi}, {Qi}, {Shi},
  {Tang}, {Tao}, {Wang}, {Wang}, {Wang}, {Wang}, {Wang}, {Wang}, {Wang},
  {Wang}, {Wang}, {Wang}, {Wang}, {Wang}, {Xu}, {Xu}, {Yang}, {Yu}, {Yuan},
  {Yuan}, {Zhai}, {Zhang}, {Zhang}, {Zhang}, {Zhao}, {Zhou}, {Zhou}, {Zhu}, \&
  {Zou}}]{2012RAA....12.1197C}
{Cui}, X.-Q., {Zhao}, Y.-H., {Chu}, Y.-Q., {et~al.} 2012, Research in Astronomy
  and Astrophysics, 12, 1197

\bibitem[{{D{\'a}lya} {et~al.}(2022){D{\'a}lya}, {D{\'\i}az}, {Bouchet},
  {Frei}, {Jasche}, {Lavaux}, {Macas}, {Mukherjee}, {P{\'a}lfi}, {de Souza},
  {Wandelt}, {Bilicki}, \& {Raffai}}]{2022MNRAS.514.1403D}
{D{\'a}lya}, G., {D{\'\i}az}, R., {Bouchet}, F.~R., {et~al.} 2022, \mnras, 514,
  1403

\bibitem[{{Drake} {et~al.}(2014){Drake}, {Graham}, {Djorgovski}, {Catelan},
  {Mahabal}, {Torrealba}, {Garc{\'\i}a-{\'A}lvarez}, {Donalek}, {Prieto},
  {Williams}, {Larson}, {Christen sen}, {Belokurov}, {Koposov}, {Beshore},
  {Boattini}, {Gibbs}, {Hill}, {Kowalski}, {Johnson}, \&
  {Shelly}}]{2014ApJS..213....9D}
{Drake}, A.~J., {Graham}, M.~J., {Djorgovski}, S.~G., {et~al.} 2014, \apjs,
  213, 9

\bibitem[{{El-Badry} {et~al.}(2021){El-Badry}, {Rix}, \&
  {Heintz}}]{2021MNRAS.506.2269E}
{El-Badry}, K., {Rix}, H.-W., \& {Heintz}, T.~M. 2021, \mnras, 506, 2269

\bibitem[{{Erkaev} {et~al.}(2007){Erkaev}, {Kulikov}, {Lammer}, {Selsis},
  {Langmayr}, {Jaritz}, \& {Biernat}}]{2007A&A...472..329E}
{Erkaev}, N.~V., {Kulikov}, Y.~N., {Lammer}, H., {et~al.} 2007, \aap, 472, 329

\bibitem[{{Favata} {et~al.}(2003){Favata}, {Giardino}, {Micela}, {Sciortino},
  \& {Damiani}}]{2003A&A...403..187F}
{Favata}, F., {Giardino}, G., {Micela}, G., {Sciortino}, S., \& {Damiani}, F.
  2003, \aap, 403, 187

\bibitem[{{Feigelson} {et~al.}(2002){Feigelson}, {Broos}, {Gaffney}, {Garmire},
  {Hillenbrand}, {Pravdo}, {Townsley}, \& {Tsuboi}}]{2002ApJ...574..258F}
{Feigelson}, E.~D., {Broos}, P., {Gaffney}, James~A., I., {et~al.} 2002, \apj,
  574, 258

\bibitem[{{Feigelson} \& {Kriss}(1981)}]{1981ApJ...248L..35F}
{Feigelson}, E.~D., \& {Kriss}, G.~A. 1981, \apjl, 248, L35

\bibitem[{{Fitzpatrick}(1999)}]{1999PASP..111...63F}
{Fitzpatrick}, E.~L. 1999, \pasp, 111, 63

\bibitem[{{Flaccomio} {et~al.}(2003){Flaccomio}, {Damiani}, {Micela},
  {Sciortino}, {Harnden}, {Murray}, \& {Wolk}}]{2003ApJ...582..398F}
{Flaccomio}, E., {Damiani}, F., {Micela}, G., {et~al.} 2003, \apj, 582, 398

\bibitem[{Foreman-Mackey(2016)}]{corner}
Foreman-Mackey, D. 2016, The Journal of Open Source Software, 1, 24.
\newblock \url{https://doi.org/10.21105/joss.00024}

\bibitem[{{Foreman-Mackey} {et~al.}(2013){Foreman-Mackey}, {Hogg}, {Lang}, \&
  {Goodman}}]{2013PASP..125..306F}
{Foreman-Mackey}, D., {Hogg}, D.~W., {Lang}, D., \& {Goodman}, J. 2013, \pasp,
  125, 306

\bibitem[{{Freund} {et~al.}(2024){Freund}, {Czesla}, {Predehl}, {Robrade},
  {Salvato}, {Schneider}, {Starck}, {Wolf}, \& {Schmitt}}]{2024arXiv240117282F}
{Freund}, S., {Czesla}, S., {Predehl}, P., {et~al.} 2024, arXiv e-prints,
  arXiv:2401.17282

\bibitem[{{Gaia Collaboration} {et~al.}(2021){Gaia Collaboration}, {Brown},
  {Vallenari}, {Prusti}, {de Bruijne}, {Babusiaux}, {Biermann}, {Creevey},
  {Evans}, {Eyer}, {Hutton}, {Jansen}, {Jordi}, {Klioner}, {Lammers},
  {Lindegren}, {Luri}, {Mignard}, {Panem}, {Pourbaix}, {Randich}, {Sartoretti},
  {Soubiran}, {Walton}, {Arenou}, {Bailer-Jones}, {Bastian}, {Cropper},
  {Drimmel}, {Katz}, {Lattanzi}, {van Leeuwen}, {Bakker}, {Cacciari},
  {Casta{\~n}eda}, {De Angeli}, {Ducourant}, {Fabricius}, {Fouesneau},
  {Fr{\'e}mat}, {Guerra}, {Guerrier}, {Guiraud}, {Jean-Antoine Piccolo},
  {Masana}, {Messineo}, {Mowlavi}, {Nicolas}, {Nienartowicz}, {Pailler},
  {Panuzzo}, {Riclet}, {Roux}, {Seabroke}, {Sordo}, {Tanga}, {Th{\'e}venin},
  {Gracia-Abril}, {Portell}, {Teyssier}, {Altmann}, {Andrae}, {Bellas-Velidis},
  {Benson}, {Berthier}, {Blomme}, {Brugaletta}, {Burgess}, {Busso}, {Carry},
  {Cellino}, {Cheek}, {Clementini}, {Damerdji}, {Davidson}, {Delchambre},
  {Dell'Oro}, {Fern{\'a}ndez-Hern{\'a}ndez}, {Galluccio}, {Garc{\'\i}a-Lario},
  {Garcia-Reinaldos}, {Gonz{\'a}lez-N{\'u}{\~n}ez}, {Gosset}, {Haigron},
  {Halbwachs}, {Hambly}, {Harrison}, {Hatzidimitriou}, {Heiter},
  {Hern{\'a}ndez}, {Hestroffer}, {Hodgkin}, {Holl}, {Jan{\ss}en}, {Jevardat de
  Fombelle}, {Jordan}, {Krone-Martins}, {Lanzafame}, {L{\"o}ffler}, {Lorca},
  {Manteiga}, {Marchal}, {Marrese}, {Moitinho}, {Mora}, {Muinonen}, {Osborne},
  {Pancino}, {Pauwels}, {Petit}, {Recio-Blanco}, {Richards}, {Riello},
  {Rimoldini}, {Robin}, {Roegiers}, {Rybizki}, {Sarro}, {Siopis}, {Smith},
  {Sozzetti}, {Ulla}, {Utrilla}, {van Leeuwen}, {van Reeven}, {Abbas}, {Abreu
  Aramburu}, {Accart}, {Aerts}, {Aguado}, {Ajaj}, {Altavilla}, {{\'A}lvarez},
  {{\'A}lvarez Cid-Fuentes}, {Alves}, {Anderson}, {Anglada Varela}, {Antoja},
  {Audard}, {Baines}, {Baker}, {Balaguer-N{\'u}{\~n}ez}, {Balbinot}, {Balog},
  {Barache}, {Barbato}, {Barros}, {Barstow}, {Bartolom{\'e}}, {Bassilana},
  {Bauchet}, {Baudesson-Stella}, {Becciani}, {Bellazzini}, {Bernet}, {Bertone},
  {Bianchi}, {Blanco-Cuaresma}, {Boch}, {Bombrun}, {Bossini}, {Bouquillon},
  {Bragaglia}, {Bramante}, {Breedt}, {Bressan}, {Brouillet}, {Bucciarelli},
  {Burlacu}, {Busonero}, {Butkevich}, {Buzzi}, {Caffau}, {Cancelliere},
  {C{\'a}novas}, {Cantat-Gaudin}, {Carballo}, {Carlucci}, {Carnerero},
  {Carrasco}, {Casamiquela}, {Castellani}, {Castro-Ginard}, {Castro Sampol},
  {Chaoul}, {Charlot}, {Chemin}, {Chiavassa}, {Cioni}, {Comoretto}, {Cooper},
  {Cornez}, {Cowell}, {Crifo}, {Crosta}, {Crowley}, {Dafonte}, {Dapergolas},
  {David}, {David}, {de Laverny}, {De Luise}, {De March}, {De Ridder}, {de
  Souza}, {de Teodoro}, {de Torres}, {del Peloso}, {del Pozo}, {Delbo},
  {Delgado}, {Delgado}, {Delisle}, {Di Matteo}, {Diakite}, {Diener},
  {Distefano}, {Dolding}, {Eappachen}, {Edvardsson}, {Enke}, {Esquej}, {Fabre},
  {Fabrizio}, {Faigler}, {Fedorets}, {Fernique}, {Fienga}, {Figueras},
  {Fouron}, {Fragkoudi}, {Fraile}, {Franke}, {Gai}, {Garabato},
  {Garcia-Gutierrez}, {Garc{\'\i}a-Torres}, {Garofalo}, {Gavras}, {Gerlach},
  {Geyer}, {Giacobbe}, {Gilmore}, {Girona}, {Giuffrida}, {Gomel}, {Gomez},
  {Gonzalez-Santamaria}, {Gonz{\'a}lez-Vidal}, {Granvik},
  {Guti{\'e}rrez-S{\'a}nchez}, {Guy}, {Hauser}, {Haywood}, {Helmi}, {Hidalgo},
  {Hilger}, {H{\l}adczuk}, {Hobbs}, {Holland}, {Huckle}, {Jasniewicz},
  {Jonker}, {Juaristi Campillo}, {Julbe}, {Karbevska}, {Kervella}, {Khanna},
  {Kochoska}, {Kontizas}, {Kordopatis}, {Korn}, {Kostrzewa-Rutkowska},
  {Kruszy{\'n}ska}, {Lambert}, {Lanza}, {Lasne}, {Le Campion}, {Le Fustec},
  {Lebreton}, {Lebzelter}, {Leccia}, {Leclerc}, {Lecoeur-Taibi}, {Liao},
  {Licata}, {Lindstr{\o}m}, {Lister}, {Livanou}, {Lobel}, {Madrero Pardo},
  {Managau}, {Mann}, {Marchant}, {Marconi}, {Marcos Santos}, {Marinoni},
  {Marocco}, {Marshall}, {Martin Polo}, {Mart{\'\i}n-Fleitas}, {Masip},
  {Massari}, {Mastrobuono-Battisti}, {Mazeh}, {McMillan}, {Messina},
  {Michalik}, {Millar}, {Mints}, {Molina}, {Molinaro}, {Moln{\'a}r},
  {Montegriffo}, {Mor}, {Morbidelli}, {Morel}, {Morris}, {Mulone}, {Munoz},
  {Muraveva}, {Murphy}, {Musella}, {Noval}, {Ord{\'e}novic}, {Orr{\`u}},
  {Osinde}, {Pagani}, {Pagano}, {Palaversa}, {Palicio}, {Panahi}, {Pawlak},
  {Pe{\~n}alosa Esteller}, {Penttil{\"a}}, {Piersimoni}, {Pineau}, {Plachy},
  {Plum}, {Poggio}, {Poretti}, {Poujoulet}, {Pr{\v{s}}a}, {Pulone}, {Racero},
  {Ragaini}, {Rainer}, {Raiteri}, {Rambaux}, {Ramos}, {Ramos-Lerate}, {Re
  Fiorentin}, {Regibo}, {Reyl{\'e}}, {Ripepi}, {Riva}, {Rixon}, {Robichon},
  {Robin}, {Roelens}, {Rohrbasser}, {Romero-G{\'o}mez}, {Rowell}, {Royer},
  {Rybicki}, {Sadowski}, {Sagrist{\`a} Sell{\'e}s}, {Sahlmann}, {Salgado},
  {Salguero}, {Samaras}, {Sanchez Gimenez}, {Sanna}, {Santove{\~n}a},
  {Sarasso}, {Schultheis}, {Sciacca}, {Segol}, {Segovia}, {S{\'e}gransan},
  {Semeux}, {Shahaf}, {Siddiqui}, {Siebert}, {Siltala}, {Slezak}, {Smart},
  {Solano}, {Solitro}, {Souami}, {Souchay}, {Spagna}, {Spoto}, {Steele},
  {Steidelm{\"u}ller}, {Stephenson}, {S{\"u}veges}, {Szabados}, {Szegedi-Elek},
  {Taris}, {Tauran}, {Taylor}, {Teixeira}, {Thuillot}, {Tonello}, {Torra},
  {Torra}, {Turon}, {Unger}, {Vaillant}, {van Dillen}, {Vanel}, {Vecchiato},
  {Viala}, {Vicente}, {Voutsinas}, {Weiler}, {Wevers}, {Wyrzykowski}, {Yoldas},
  {Yvard}, {Zhao}, {Zorec}, {Zucker}, {Zurbach}, \&
  {Zwitter}}]{2021A&A...649A...1G}
{Gaia Collaboration}, {Brown}, A.~G.~A., {Vallenari}, A., {et~al.} 2021, \aap,
  649, A1

\bibitem[{{Green}(2018)}]{2018JOSS....3..695M}
{Green}, G. 2018, The Journal of Open Source Software, 3, 695

\bibitem[{{Green} {et~al.}(2019){Green}, {Schlafly}, {Zucker}, {Speagle}, \&
  {Finkbeiner}}]{2019ApJ...887...93G}
{Green}, G.~M., {Schlafly}, E., {Zucker}, C., {Speagle}, J.~S., \&
  {Finkbeiner}, D. 2019, \apj, 887, 93

\bibitem[{{G{\"u}del} {et~al.}(1997){G{\"u}del}, {Guinan}, \&
  {Skinner}}]{1997ApJ...483..947G}
{G{\"u}del}, M., {Guinan}, E.~F., \& {Skinner}, S.~L. 1997, \apj, 483, 947

\bibitem[{{Han} {et~al.}(2023){Han}, {Wang}, {Bai}, {Yang}, {Fang}, \&
  {Liu}}]{2023ApJS..264...12H}
{Han}, H., {Wang}, S., {Bai}, Y., {et~al.} 2023, \apjs, 264, 12

\bibitem[{{Hornschemeier} {et~al.}(2003){Hornschemeier}, {Bauer}, {Alexander},
  {Brandt}, {Sargent}, {Bautz}, {Conselice}, {Garmire}, {Schneider}, \&
  {Wilson}}]{2003AJ....126..575H}
{Hornschemeier}, A.~E., {Bauer}, F.~E., {Alexander}, D.~M., {et~al.} 2003, \aj,
  126, 575

\bibitem[{{Howard} {et~al.}(2022){Howard}, {Davenport}, \&
  {Covey}}]{2022RNAAS...6...96H}
{Howard}, E.~L., {Davenport}, J. R.~A., \& {Covey}, K.~R. 2022, Research Notes
  of the American Astronomical Society, 6, 96

\bibitem[{{James} {et~al.}(2000){James}, {Jardine}, {Jeffries}, {Randich},
  {Collier Cameron}, \& {Ferreira}}]{2000MNRAS.318.1217J}
{James}, D.~J., {Jardine}, M.~M., {Jeffries}, R.~D., {et~al.} 2000, \mnras,
  318, 1217

\bibitem[{{Jayasinghe} {et~al.}(2020){Jayasinghe}, {Stanek}, {Kochanek},
  {Shappee}, {Holoien}, {Thompson}, {Prieto}, {Dong}, {Pawlak}, {Pejcha},
  {Shields}, {Pojmanski}, {Otero}, {Hurst}, {Britt}, \&
  {Will}}]{2020MNRAS.491...13J}
{Jayasinghe}, T., {Stanek}, K.~Z., {Kochanek}, C.~S., {et~al.} 2020, \mnras,
  491, 13

\bibitem[{{Jenkins} {et~al.}(2016){Jenkins}, {Twicken}, {McCauliff},
  {Campbell}, {Sanderfer}, {Lung}, {Mansouri-Samani}, {Girouard}, {Tenenbaum},
  {Klaus}, {Smith}, {Caldwell}, {Chacon}, {Henze}, {Heiges}, {Latham},
  {Morgan}, {Swade}, {Rinehart}, \& {Vanderspek}}]{2016SPIE.9913E..3EJ}
{Jenkins}, J.~M., {Twicken}, J.~D., {McCauliff}, S., {et~al.} 2016, in Society
  of Photo-Optical Instrumentation Engineers (SPIE) Conference Series, Vol.
  9913, Software and Cyberinfrastructure for Astronomy IV, ed. G.~{Chiozzi} \&
  J.~C. {Guzman}, 99133E

\bibitem[{{Johnstone} {et~al.}(2021){Johnstone}, {Bartel}, \&
  {G{\"u}del}}]{2021A&A...649A..96J}
{Johnstone}, C.~P., {Bartel}, M., \& {G{\"u}del}, M. 2021, \aap, 649, A96

\bibitem[{{J{\"o}nsson} {et~al.}(2020){J{\"o}nsson}, {Holtzman}, {Allende
  Prieto}, {Cunha}, {Garc{\'\i}a-Hern{\'a}ndez}, {Hasselquist}, {Masseron},
  {Osorio}, {Shetrone}, {Smith}, {Stringfellow}, {Bizyaev}, {Edvardsson},
  {Majewski}, {M{\'e}sz{\'a}ros}, {Souto}, {Zamora}, {Beaton}, {Bovy}, {Donor},
  {Pinsonneault}, {Poovelil}, \& {Sobeck}}]{2020AJ....160..120J}
{J{\"o}nsson}, H., {Holtzman}, J.~A., {Allende Prieto}, C., {et~al.} 2020, \aj,
  160, 120

\bibitem[{{Kasting} {et~al.}(1993){Kasting}, {Whitmire}, \&
  {Reynolds}}]{1993Icar..101..108K}
{Kasting}, J.~F., {Whitmire}, D.~P., \& {Reynolds}, R.~T. 1993, \icarus, 101,
  108

\bibitem[{{Kite} \& {Barnett}(2020)}]{2020PNAS..11718264K}
{Kite}, E.~S., \& {Barnett}, M.~N. 2020, Proceedings of the National Academy of
  Science, 117, 18264

\bibitem[{{Kopparapu} {et~al.}(2014){Kopparapu}, {Ramirez}, {SchottelKotte},
  {Kasting}, {Domagal-Goldman}, \& {Eymet}}]{2014ApJ...787L..29K}
{Kopparapu}, R.~K., {Ramirez}, R.~M., {SchottelKotte}, J., {et~al.} 2014,
  \apjl, 787, L29

\bibitem[{{Kopparapu} {et~al.}(2013){Kopparapu}, {Ramirez}, {Kasting}, {Eymet},
  {Robinson}, {Mahadevan}, {Terrien}, {Domagal-Goldman}, {Meadows}, \&
  {Deshpande}}]{2013ApJ...765..131K}
{Kopparapu}, R.~K., {Ramirez}, R., {Kasting}, J.~F., {et~al.} 2013, \apj, 765,
  131

\bibitem[{{Kounkel} {et~al.}(2021){Kounkel}, {Covey}, {Stassun},
  {Price-Whelan}, {Holtzman}, {Chojnowski}, {Longa-Pe{\~n}a},
  {Rom{\'a}n-Z{\'u}{\~n}iga}, {Hernandez}, {Serna}, {Badenes}, {De Lee},
  {Majewski}, {Stringfellow}, {Kratter}, {Moe}, {Frinchaboy}, {Beaton},
  {Fern{\'a}ndez-Trincado}, {Mahadevan}, {Minniti}, {Beers}, {Schneider},
  {Barba}, {Brownstein}, {Garc{\'\i}a-Hern{\'a}ndez}, {Pan}, \&
  {Bizyaev}}]{2021AJ....162..184K}
{Kounkel}, M., {Covey}, K.~R., {Stassun}, K.~G., {et~al.} 2021, \aj, 162, 184

\bibitem[{{Kovalev} {et~al.}(2022){Kovalev}, {Chen}, \&
  {Han}}]{2022MNRAS.517..356K}
{Kovalev}, M., {Chen}, X., \& {Han}, Z. 2022, \mnras, 517, 356

\bibitem[{{Kubyshkina} {et~al.}(2020){Kubyshkina}, {Vidotto}, {Fossati}, \&
  {Farrell}}]{2020MNRAS.499...77K}
{Kubyshkina}, D., {Vidotto}, A.~A., {Fossati}, L., \& {Farrell}, E. 2020,
  \mnras, 499, 77

\bibitem[{{Lammer} {et~al.}(2009){Lammer}, {Odert}, {Leitzinger},
  {Khodachenko}, {Panchenko}, {Kulikov}, {Zhang}, {Lichtenegger}, {Erkaev},
  {Wuchterl}, {Micela}, {Penz}, {Biernat}, {Weingrill}, {Steller}, {Ottacher},
  {Hasiba}, \& {Hanslmeier}}]{2009A&A...506..399L}
{Lammer}, H., {Odert}, P., {Leitzinger}, M., {et~al.} 2009, \aap, 506, 399

\bibitem[{{Lehtinen} {et~al.}(2020){Lehtinen}, {Spada}, {K{\"a}pyl{\"a}},
  {Olspert}, \& {K{\"a}pyl{\"a}}}]{2020NatAs...4..658L}
{Lehtinen}, J.~J., {Spada}, F., {K{\"a}pyl{\"a}}, M.~J., {Olspert}, N., \&
  {K{\"a}pyl{\"a}}, P.~J. 2020, Nature Astronomy, 4, 658

\bibitem[{{Li} {et~al.}(2021){Li}, {Shi}, {Yan}, {Fu}, {Li}, \&
  {Hou}}]{2021ApJS..256...31L}
{Li}, C.-q., {Shi}, J.-r., {Yan}, H.-l., {et~al.} 2021, \apjs, 256, 31

\bibitem[{{Lomb}(1976)}]{1976Ap&SS..39..447L}
{Lomb}, N.~R. 1976, \apss, 39, 447

\bibitem[{{Luo} {et~al.}(2015){Luo}, {Zhao}, {Zhao}, {Deng}, {Liu}, {Jing},
  {Wang}, {Zhang}, {Shi}, {Cui}, {Chu}, {Li}, {Bai}, {Wu}, {Cai}, {Cao}, {Cao},
  {Carlin}, {Chen}, {Chen}, {Chen}, {Chen}, {Chen}, {Chen}, {Chen},
  {Christlieb}, {Chu}, {Cui}, {Dong}, {Du}, {Fan}, {Feng}, {Fu}, {Gao}, {Gong},
  {Gu}, {Guo}, {Han}, {He}, {Hou}, {Hou}, {Hou}, {Hu}, {Hu}, {Hu}, {Huo},
  {Jia}, {Jiang}, {Jiang}, {Jiang}, {Jin}, {Kong}, {Kong}, {Lei}, {Li}, {Li},
  {Li}, {Li}, {Li}, {Li}, {Li}, {Li}, {Li}, {Li}, {Li}, {Li}, {Liang}, {Lin},
  {Liu}, {Liu}, {Liu}, {Liu}, {Lu}, {Luo}, {Mao}, {Newberg}, {Ni}, {Qi}, {Qi},
  {Shen}, {Shi}, {Song}, {Song}, {Su}, {Su}, {Tang}, {Tao}, {Tian}, {Wang},
  {Wang}, {Wang}, {Wang}, {Wang}, {Wang}, {Wang}, {Wang}, {Wang}, {Wang},
  {Wang}, {Wang}, {Wang}, {Wang}, {Wang}, {Wang}, {Wang}, {Wang}, {Wang},
  {Wang}, {Wei}, {Wei}, {Wu}, {Wu}, {Wu}, {Wu}, {Xing}, {Xu}, {Xu}, {Xu},
  {Yan}, {Yang}, {Yang}, {Yang}, {Yang}, {Yao}, {Yu}, {Yuan}, {Yuan}, {Yuan},
  {Yuan}, {Zhai}, {Zhang}, {Zhang}, {Zhang}, {Zhang}, {Zhang}, {Zhang},
  {Zhang}, {Zhang}, {Zhao}, {Zhou}, {Zhou}, {Zhu}, {Zhu}, {Zou}, \&
  {Zuo}}]{2015RAA....15.1095L}
{Luo}, A.~L., {Zhao}, Y.-H., {Zhao}, G., {et~al.} 2015, Research in Astronomy
  and Astrophysics, 15, 1095

\bibitem[{{Mamajek} \& {Hillenbrand}(2008)}]{2008ApJ...687.1264M}
{Mamajek}, E.~E., \& {Hillenbrand}, L.~A. 2008, \apj, 687, 1264

\bibitem[{{Marton} {et~al.}(2016){Marton}, {T{\'o}th}, {Paladini}, {Kun},
  {Zahorecz}, {McGehee}, \& {Kiss}}]{2016MNRAS.458.3479M}
{Marton}, G., {T{\'o}th}, L.~V., {Paladini}, R., {et~al.} 2016, \mnras, 458,
  3479

\bibitem[{{Marton} {et~al.}(2019){Marton}, {{\'A}brah{\'a}m}, {Szegedi-Elek},
  {Varga}, {Kun}, {K{\'o}sp{\'a}l}, {Varga-Vereb{\'e}lyi}, {Hodgkin},
  {Szabados}, {Beck}, \& {Kiss}}]{2019MNRAS.487.2522M}
{Marton}, G., {{\'A}brah{\'a}m}, P., {Szegedi-Elek}, E., {et~al.} 2019, \mnras,
  487, 2522

\bibitem[{{Marton} {et~al.}(2023){Marton}, {{\'A}brah{\'a}m}, {Rimoldini},
  {Audard}, {Kun}, {Nagy}, {K{\'o}sp{\'a}l}, {Szabados}, {Holl}, {Gavras},
  {Mowlavi}, {Nienartowicz}, {de Fombelle}, {Lecoeur-Ta{\"\i}bi}, {Karbevska},
  {Lario}, \& {Eyer}}]{2023A&A...674A..21M}
{Marton}, G., {{\'A}brah{\'a}m}, P., {Rimoldini}, L., {et~al.} 2023, \aap, 674,
  A21

\bibitem[{{MAST Team}(2021)}]{https://doi.org/10.17909/t9-nmc8-f686}
{MAST Team}. 2021, TESS Light Curves - All Sectors,  STScI/MAST,
  doi:10.17909/T9-NMC8-F686

\bibitem[{{McQuillan} {et~al.}(2013){McQuillan}, {Aigrain}, \&
  {Mazeh}}]{2013MNRAS.432.1203M}
{McQuillan}, A., {Aigrain}, S., \& {Mazeh}, T. 2013, \mnras, 432, 1203

\bibitem[{{McQuillan} {et~al.}(2014){McQuillan}, {Mazeh}, \&
  {Aigrain}}]{2014ApJS..211...24M}
{McQuillan}, A., {Mazeh}, T., \& {Aigrain}, S. 2014, \apjs, 211, 24

\bibitem[{{Mittag} {et~al.}(2018){Mittag}, {Schmitt}, \&
  {Schr{\"o}der}}]{2018A&A...618A..48M}
{Mittag}, M., {Schmitt}, J.~H.~M.~M., \& {Schr{\"o}der}, K.~P. 2018, \aap, 618,
  A48

\bibitem[{{Morton}(2015)}]{2015ascl.soft03010M}
{Morton}, T.~D. 2015, {isochrones: Stellar model grid package}, Astrophysics
  Source Code Library, record ascl:1503.010, , , ascl:1503.010

\bibitem[{{Motch} {et~al.}(2010){Motch}, {Warwick}, {Cropper}, {Carrera},
  {Guillout}, {Pineau}, {Pakull}, {Rosen}, {Schwope}, {Tedds}, {Webb},
  {Negueruela}, \& {Watson}}]{2010A&A...523A..92M}
{Motch}, C., {Warwick}, R., {Cropper}, M.~S., {et~al.} 2010, \aap, 523, A92

\bibitem[{{Murray-Clay} {et~al.}(2009){Murray-Clay}, {Chiang}, \&
  {Murray}}]{2009ApJ...693...23M}
{Murray-Clay}, R.~A., {Chiang}, E.~I., \& {Murray}, N. 2009, \apj, 693, 23

\bibitem[{{Newton} {et~al.}(2017){Newton}, {Irwin}, {Charbonneau}, {Berlind},
  {Calkins}, \& {Mink}}]{2017ApJ...834...85N}
{Newton}, E.~R., {Irwin}, J., {Charbonneau}, D., {et~al.} 2017, \apj, 834, 85

\bibitem[{{Noyes} {et~al.}(1984){Noyes}, {Hartmann}, {Baliunas}, {Duncan}, \&
  {Vaughan}}]{1984ApJ...279..763N}
{Noyes}, R.~W., {Hartmann}, L.~W., {Baliunas}, S.~L., {Duncan}, D.~K., \&
  {Vaughan}, A.~H. 1984, \apj, 279, 763

\bibitem[{{Owen} \& {Mohanty}(2016)}]{2016MNRAS.459.4088O}
{Owen}, J.~E., \& {Mohanty}, S. 2016, \mnras, 459, 4088

\bibitem[{{Pallavicini} {et~al.}(1981){Pallavicini}, {Golub}, {Rosner},
  {Vaiana}, {Ayres}, \& {Linsky}}]{1981ApJ...248..279P}
{Pallavicini}, R., {Golub}, L., {Rosner}, R., {et~al.} 1981, \apj, 248, 279

\bibitem[{{Parker}(1955)}]{1955ApJ...122..293P}
{Parker}, E.~N. 1955, \apj, 122, 293

\bibitem[{{Parker}(1988)}]{1988ApJ...330..474P}
---. 1988, \apj, 330, 474

\bibitem[{{Pizzocaro} {et~al.}(2019){Pizzocaro}, {Stelzer}, {Poretti}, {Raetz},
  {Micela}, {Belfiore}, {Marelli}, {Salvetti}, \& {De
  Luca}}]{2019A&A...628A..41P}
{Pizzocaro}, D., {Stelzer}, B., {Poretti}, E., {et~al.} 2019, \aap, 628, A41

\bibitem[{{Pizzolato} {et~al.}(2003){Pizzolato}, {Maggio}, {Micela},
  {Sciortino}, \& {Ventura}}]{2003A&A...397..147P}
{Pizzolato}, N., {Maggio}, A., {Micela}, G., {Sciortino}, S., \& {Ventura}, P.
  2003, \aap, 397, 147

\bibitem[{{Powner} {et~al.}(2009){Powner}, {Gerland}, \&
  {Sutherland}}]{2009Natur.459..239P}
{Powner}, M.~W., {Gerland}, B., \& {Sutherland}, J.~D. 2009, \nat, 459, 239

\bibitem[{{Preibisch} \& {Feigelson}(2005)}]{2005ApJS..160..390P}
{Preibisch}, T., \& {Feigelson}, E.~D. 2005, \apjs, 160, 390

\bibitem[{{Pr{\v{s}}a} {et~al.}(2022){Pr{\v{s}}a}, {Kochoska}, {Conroy},
  {Eisner}, {Hey}, {IJspeert}, {Kruse}, {Fleming}, {Johnston}, {Kristiansen},
  {LaCourse}, {Mortensen}, {Pepper}, {Stassun}, {Torres}, {Abdul-Masih},
  {Chakraborty}, {Gagliano}, {Guo}, {Hambleton}, {Hong}, {Jacobs}, {Jones},
  {Kostov}, {Lee}, {Omohundro}, {Orosz}, {Page}, {Powell}, {Rappaport}, {Reed},
  {Schnittman}, {Schwengeler}, {Shporer}, {Terentev}, {Vanderburg}, {Welsh},
  {Caldwell}, {Doty}, {Jenkins}, {Latham}, {Ricker}, {Seager}, {Schlieder},
  {Shiao}, {Vanderspek}, \& {Winn}}]{2022ApJS..258...16P}
{Pr{\v{s}}a}, A., {Kochoska}, A., {Conroy}, K.~E., {et~al.} 2022, \apjs, 258,
  16

\bibitem[{{Randich} {et~al.}(1996){Randich}, {Schmitt}, {Prosser}, \&
  {Stauffer}}]{1996A&A...305..785R}
{Randich}, S., {Schmitt}, J.~H.~M.~M., {Prosser}, C.~F., \& {Stauffer}, J.~R.
  1996, \aap, 305, 785

\bibitem[{{Reiners} {et~al.}(2014){Reiners}, {Sch{\"u}ssler}, \&
  {Passegger}}]{2014ApJ...794..144R}
{Reiners}, A., {Sch{\"u}ssler}, M., \& {Passegger}, V.~M. 2014, \apj, 794, 144

\bibitem[{{Ren} {et~al.}(2018){Ren}, {Rebassa-Mansergas}, {Parsons}, {Liu},
  {Luo}, {Kong}, \& {Zhang}}]{2018MNRAS.477.4641R}
{Ren}, J.~J., {Rebassa-Mansergas}, A., {Parsons}, S.~G., {et~al.} 2018, \mnras,
  477, 4641

\bibitem[{{Ren} {et~al.}(2020){Ren}, {Raddi}, {Rebassa-Mansergas}, {Hernandez},
  {Parsons}, {Irawati}, {Rittipruk}, {Schreiber}, {G{\"a}nsicke}, {Torres},
  {Wang}, {Zhang}, {Zhao}, {Zhou}, {Han}, {Wang}, {Liu}, {Liu}, {Wang},
  {Zheng}, {Wang}, {Zhao}, {Cui}, {Shi}, \& {Tian}}]{2020ApJ...905...38R}
{Ren}, J.~J., {Raddi}, R., {Rebassa-Mansergas}, A., {et~al.} 2020, \apj, 905,
  38

\bibitem[{{Richey-Yowell} {et~al.}(2023){Richey-Yowell}, {Shkolnik},
  {Schneider}, {Peacock}, {Huseby}, {Jackman}, {Barman}, {Osby}, \&
  {Meadows}}]{2023ApJ...951...44R}
{Richey-Yowell}, T., {Shkolnik}, E.~L., {Schneider}, A.~C., {et~al.} 2023,
  \apj, 951, 44

\bibitem[{{Ricker} {et~al.}(2015){Ricker}, {Winn}, {Vanderspek}, {Latham},
  {Bakos}, {Bean}, {Berta-Thompson}, {Brown}, {Buchhave}, {Butler}, {Butler},
  {Chaplin}, {Charbonneau}, {Christensen-Dalsgaard}, {Clampin}, {Deming},
  {Doty}, {De Lee}, {Dressing}, {Dunham}, {Endl}, {Fressin}, {Ge}, {Henning},
  {Holman}, {Howard}, {Ida}, {Jenkins}, {Jernigan}, {Johnson}, {Kaltenegger},
  {Kawai}, {Kjeldsen}, {Laughlin}, {Levine}, {Lin}, {Lissauer}, {MacQueen},
  {Marcy}, {McCullough}, {Morton}, {Narita}, {Paegert}, {Palle}, {Pepe},
  {Pepper}, {Quirrenbach}, {Rinehart}, {Sasselov}, {Sato}, {Seager},
  {Sozzetti}, {Stassun}, {Sullivan}, {Szentgyorgyi}, {Torres}, {Udry}, \&
  {Villasenor}}]{2015JATIS...1a4003R}
{Ricker}, G.~R., {Winn}, J.~N., {Vanderspek}, R., {et~al.} 2015, Journal of
  Astronomical Telescopes, Instruments, and Systems, 1, 014003

\bibitem[{{Rimoldini} {et~al.}(2023){Rimoldini}, {Holl}, {Gavras}, {Audard},
  {De Ridder}, {Mowlavi}, {Nienartowicz}, {Jevardat de Fombelle},
  {Lecoeur-Ta{\"\i}bi}, {Karbevska}, {Evans}, {{\'A}brah{\'a}m}, {Carnerero},
  {Clementini}, {Distefano}, {Garofalo}, {Garc{\'\i}a-Lario}, {Gomel},
  {Klioner}, {Kruszy{\'n}ska}, {Lanzafame}, {Lebzelter}, {Marton}, {Mazeh},
  {Molinaro}, {Panahi}, {Raiteri}, {Ripepi}, {Szabados}, {Teyssier},
  {Trabucchi}, {Wyrzykowski}, {Zucker}, \& {Eyer}}]{2023A&A...674A..14R}
{Rimoldini}, L., {Holl}, B., {Gavras}, P., {et~al.} 2023, \aap, 674, A14

\bibitem[{{Rogers} {et~al.}(2011){Rogers}, {Bodenheimer}, {Lissauer}, \&
  {Seager}}]{2011ApJ...738...59R}
{Rogers}, L.~A., {Bodenheimer}, P., {Lissauer}, J.~J., \& {Seager}, S. 2011,
  \apj, 738, 59

\bibitem[{{Salvato} {et~al.}(2018){Salvato}, {Buchner}, {Budav{\'a}ri},
  {Dwelly}, {Merloni}, {Brusa}, {Rau}, {Fotopoulou}, \&
  {Nandra}}]{2018MNRAS.473.4937S}
{Salvato}, M., {Buchner}, J., {Budav{\'a}ri}, T., {et~al.} 2018, \mnras, 473,
  4937

\bibitem[{{Samus'} {et~al.}(2017){Samus'}, {Kazarovets}, {Durlevich},
  {Kireeva}, \& {Pastukhova}}]{2017ARep...61...80S}
{Samus'}, N.~N., {Kazarovets}, E.~V., {Durlevich}, O.~V., {Kireeva}, N.~N., \&
  {Pastukhova}, E.~N. 2017, Astronomy Reports, 61, 80

\bibitem[{{Santos} {et~al.}(2021){Santos}, {Breton}, {Mathur}, \&
  {Garc{\'\i}a}}]{2021ApJS..255...17S}
{Santos}, A.~R.~G., {Breton}, S.~N., {Mathur}, S., \& {Garc{\'\i}a}, R.~A.
  2021, \apjs, 255, 17

\bibitem[{{Santos} {et~al.}(2019){Santos}, {Garc{\'\i}a}, {Mathur}, {Bugnet},
  {van Saders}, {Metcalfe}, {Simonian}, \&
  {Pinsonneault}}]{2019ApJS..244...21S}
{Santos}, A.~R.~G., {Garc{\'\i}a}, R.~A., {Mathur}, S., {et~al.} 2019, \apjs,
  244, 21

\bibitem[{{Scargle}(1982)}]{1982ApJ...263..835S}
{Scargle}, J.~D. 1982, \apj, 263, 835

\bibitem[{{Schlegel} {et~al.}(1998){Schlegel}, {Finkbeiner}, \&
  {Davis}}]{1998ApJ...500..525S}
{Schlegel}, D.~J., {Finkbeiner}, D.~P., \& {Davis}, M. 1998, \apj, 500, 525

\bibitem[{{Segura} {et~al.}(2003){Segura}, {Krelove}, {Kasting}, {Sommerlatt},
  {Meadows}, {Crisp}, {Cohen}, \& {Mlawer}}]{2003AsBio...3..689S}
{Segura}, A., {Krelove}, K., {Kasting}, J.~F., {et~al.} 2003, Astrobiology, 3,
  689

\bibitem[{{Soszy{\'n}ski} {et~al.}(2016){Soszy{\'n}ski}, {Pawlak},
  {Pietrukowicz}, {Udalski}, {Szyma{\'n}ski}, {Wyrzykowski}, {Ulaczyk},
  {Poleski}, {Koz{\l}owski}, {Skowron}, {Skowron}, {Mr{\'o}z}, \&
  {Hamanowicz}}]{2016AcA....66..405S}
{Soszy{\'n}ski}, I., {Pawlak}, M., {Pietrukowicz}, P., {et~al.} 2016, \actaa,
  66, 405

\bibitem[{{Spada} {et~al.}(2017){Spada}, {Demarque}, {Kim}, {Boyajian}, \&
  {Brewer}}]{2017ApJ...838..161S}
{Spada}, F., {Demarque}, P., {Kim}, Y.~C., {Boyajian}, T.~S., \& {Brewer},
  J.~M. 2017, \apj, 838, 161

\bibitem[{{Spinelli} {et~al.}(2023){Spinelli}, {Borsa}, {Ghirlanda},
  {Ghisellini}, \& {Haardt}}]{2023MNRAS.522.1411S}
{Spinelli}, R., {Borsa}, F., {Ghirlanda}, G., {Ghisellini}, G., \& {Haardt}, F.
  2023, \mnras, 522, 1411

\bibitem[{{Stassun} {et~al.}(2019){Stassun}, {Oelkers}, {Paegert}, {Torres},
  {Pepper}, {De Lee}, {Collins}, {Latham}, {Muirhead}, {Chittidi},
  {Rojas-Ayala}, {Fleming}, {Rose}, {Tenenbaum}, {Ting}, {Kane}, {Barclay},
  {Bean}, {Brassuer}, {Charbonneau}, {Ge}, {Lissauer}, {Mann}, {McLean},
  {Mullally}, {Narita}, {Plavchan}, {Ricker}, {Sasselov}, {Seager}, {Sharma},
  {Shiao}, {Sozzetti}, {Stello}, {Vanderspek}, {Wallace}, \&
  {Winn}}]{2019AJ....158..138S}
{Stassun}, K.~G., {Oelkers}, R.~J., {Paegert}, M., {et~al.} 2019, \aj, 158, 138

\bibitem[{{Stauffer} {et~al.}(1997){Stauffer}, {Hartmann}, {Prosser},
  {Randich}, {Balachandran}, {Patten}, {Simon}, \&
  {Giampapa}}]{1997ApJ...479..776S}
{Stauffer}, J.~R., {Hartmann}, L.~W., {Prosser}, C.~F., {et~al.} 1997, \apj,
  479, 776

\bibitem[{{St{\c{e}}pie{\'n}} {et~al.}(2001){St{\c{e}}pie{\'n}}, {Schmitt}, \&
  {Voges}}]{2001A&A...370..157S}
{St{\c{e}}pie{\'n}}, K., {Schmitt}, J.~H.~M.~M., \& {Voges}, W. 2001, \aap,
  370, 157

\bibitem[{{Stocke} {et~al.}(1991){Stocke}, {Morris}, {Gioia}, {Maccacaro},
  {Schild}, {Wolter}, {Fleming}, \& {Henry}}]{1991ApJS...76..813S}
{Stocke}, J.~T., {Morris}, S.~L., {Gioia}, I.~M., {et~al.} 1991, \apjs, 76, 813

\bibitem[{{Swain} {et~al.}(2021){Swain}, {Estrela}, {Roudier}, {Sotin},
  {Rimmer}, {Valio}, {West}, {Pearson}, {Huber-Feely}, \&
  {Zellem}}]{2021AJ....161..213S}
{Swain}, M.~R., {Estrela}, R., {Roudier}, G.~M., {et~al.} 2021, \aj, 161, 213

\bibitem[{{Truemper}(1982)}]{1982AdSpR...2d.241T}
{Truemper}, J. 1982, Advances in Space Research, 2, 241

\bibitem[{{Vaiana} {et~al.}(1981){Vaiana}, {Cassinelli}, {Fabbiano},
  {Giacconi}, {Golub}, {Gorenstein}, {Haisch}, {Harnden}, {Johnson}, {Linsky},
  {Maxson}, {Mewe}, {Rosner}, {Seward}, {Topka}, \&
  {Zwaan}}]{1981ApJ...245..163V}
{Vaiana}, G.~S., {Cassinelli}, J.~P., {Fabbiano}, G., {et~al.} 1981, \apj, 245,
  163

\bibitem[{{van Ballegooijen} {et~al.}(2011){van Ballegooijen}, {Asgari-Targhi},
  {Cranmer}, \& {DeLuca}}]{2011ApJ...736....3V}
{van Ballegooijen}, A.~A., {Asgari-Targhi}, M., {Cranmer}, S.~R., \& {DeLuca},
  E.~E. 2011, \apj, 736, 3

\bibitem[{{Voges} {et~al.}(1999){Voges}, {Aschenbach}, {Boller},
  {Br{\"a}uninger}, {Briel}, {Burkert}, {Dennerl}, {Englhauser}, {Gruber},
  {Haberl}, {Hartner}, {Hasinger}, {K{\"u}rster}, {Pfeffermann}, {Pietsch},
  {Predehl}, {Rosso}, {Schmitt}, {Tr{\"u}mper}, \&
  {Zimmermann}}]{1999A&A...349..389V}
{Voges}, W., {Aschenbach}, B., {Boller}, T., {et~al.} 1999, \aap, 349, 389

\bibitem[{{Wang} {et~al.}(2020){Wang}, {Bai}, {He}, \&
  {Liu}}]{2020ApJ...902..114W}
{Wang}, S., {Bai}, Y., {He}, L., \& {Liu}, J. 2020, \apj, 902, 114

\bibitem[{{Wang} {et~al.}(2016){Wang}, {Liu}, {Qiu}, {Bai}, {Yang}, {Guo}, \&
  {Zhang}}]{2016ApJS..224...40W}
{Wang}, S., {Liu}, J., {Qiu}, Y., {et~al.} 2016, \apjs, 224, 40

\bibitem[{{Wei}(2022)}]{2022ApJ...926...40W}
{Wei}, X. 2022, \apj, 926, 40

\bibitem[{{Wolfgang} \& {Lopez}(2015)}]{2015ApJ...806..183W}
{Wolfgang}, A., \& {Lopez}, E. 2015, \apj, 806, 183

\bibitem[{{Wright} {et~al.}(2011){Wright}, {Drake}, {Mamajek}, \&
  {Henry}}]{2011ApJ...743...48W}
{Wright}, N.~J., {Drake}, J.~J., {Mamajek}, E.~E., \& {Henry}, G.~W. 2011,
  \apj, 743, 48

\bibitem[{{Wright} {et~al.}(2018){Wright}, {Newton}, {Williams}, {Drake}, \&
  {Yadav}}]{2018MNRAS.479.2351W}
{Wright}, N.~J., {Newton}, E.~R., {Williams}, P. K.~G., {Drake}, J.~J., \&
  {Yadav}, R.~K. 2018, \mnras, 479, 2351

\bibitem[{{Zhang} {et~al.}(2022){Zhang}, {Jing}, {Yang}, {Wan}, {Ji}, {Fu},
  {Liu}, {Zhang}, {Luo}, {Tian}, {Zhou}, {Wang}, {Guo}, {Zong}, {Xiong}, \&
  {Li}}]{2022ApJS..258...26Z}
{Zhang}, B., {Jing}, Y.-J., {Yang}, F., {et~al.} 2022, \apjs, 258, 26

\bibitem[{{Zhu} {et~al.}(2017){Zhu}, {Tian}, {Li}, \&
  {Zhang}}]{2017MNRAS.471.3494Z}
{Zhu}, H., {Tian}, W., {Li}, A., \& {Zhang}, M. 2017, \mnras, 471, 3494

\end{thebibliography}

\clearpage
\begin{appendix}

\section{$R_X$ comparisons between this work and previous studies}
\renewcommand\thefigure{\Alph{section}\arabic{figure}}
\setcounter{figure}{0}

\begin{figure}[h]
\centering
\includegraphics[width=0.5\textwidth]{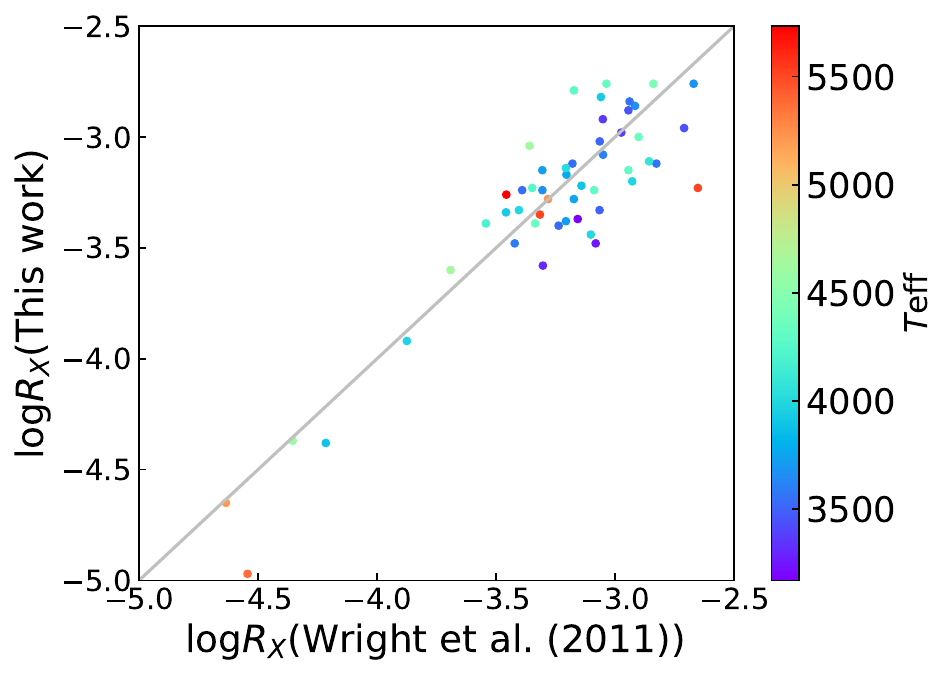}
\caption{Comparison of $R_X$ of common targets between this work and \cite{2011ApJ...743...48W}. Different colors represent different effective temperatures.}
\label{cross_wright.fig}
\end{figure}

\begin{figure}[h]
\centering
\includegraphics[width=0.45\textwidth]{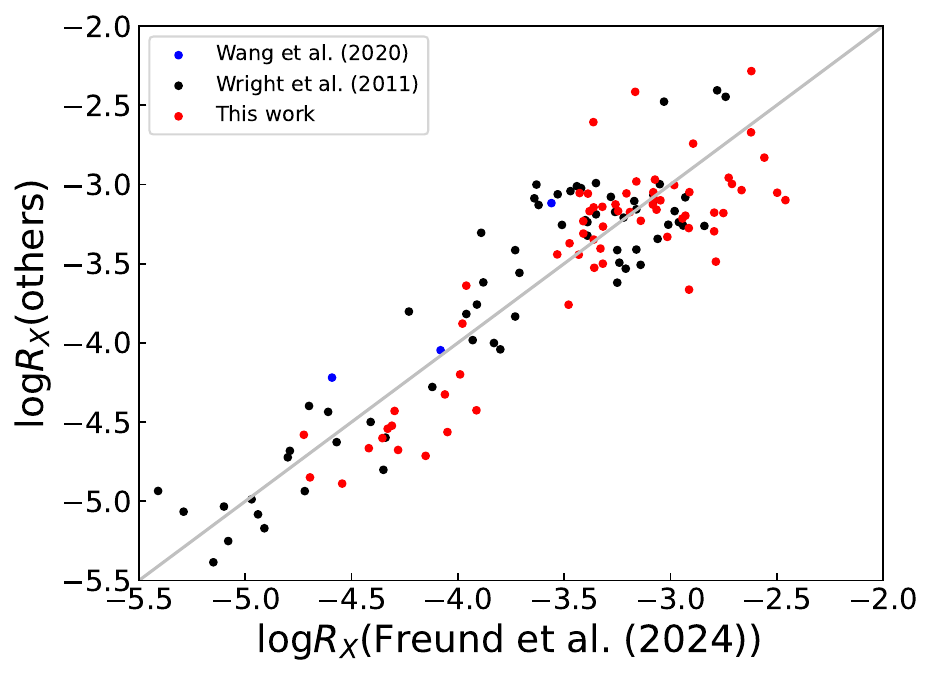}
\caption{Comparison of $R_X$ of common targets between various literature and \cite{2024arXiv240117282F}. Different colors represent sources from different literature.}
\label{cross_various.fig}
\end{figure}

Figure \ref{cross_wright.fig} shows the comparison of $R_X$ between this work and \cite{2011ApJ...743...48W}. Figure \ref{cross_various.fig} shows the comparison of $R_X$ between \cite{2024arXiv240117282F} and other studies. 

\clearpage
\section{MCMC results of the fits to activity-rotation relations}
Figure \ref{MCMC} shows the posterior distributions of parameters of the best-fit piece-wise power laws. 

\setcounter{figure}{0}
\begin{figure*}[h]
%\centering
\subfigure[]{
\includegraphics[width=0.5\textwidth]{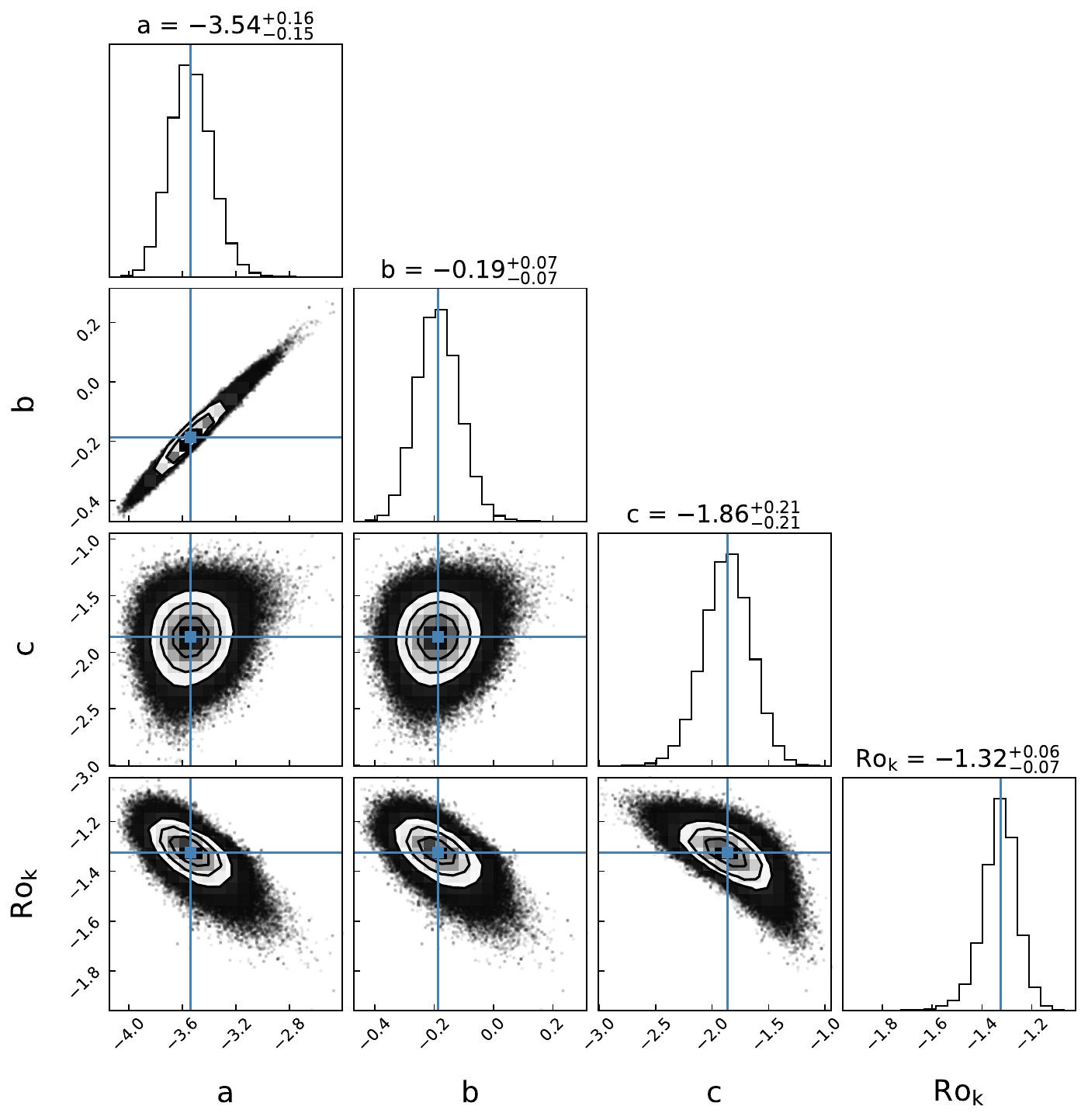}}
%\label{HRD.fig}
\subfigure[]{
\includegraphics[width=0.5\textwidth]{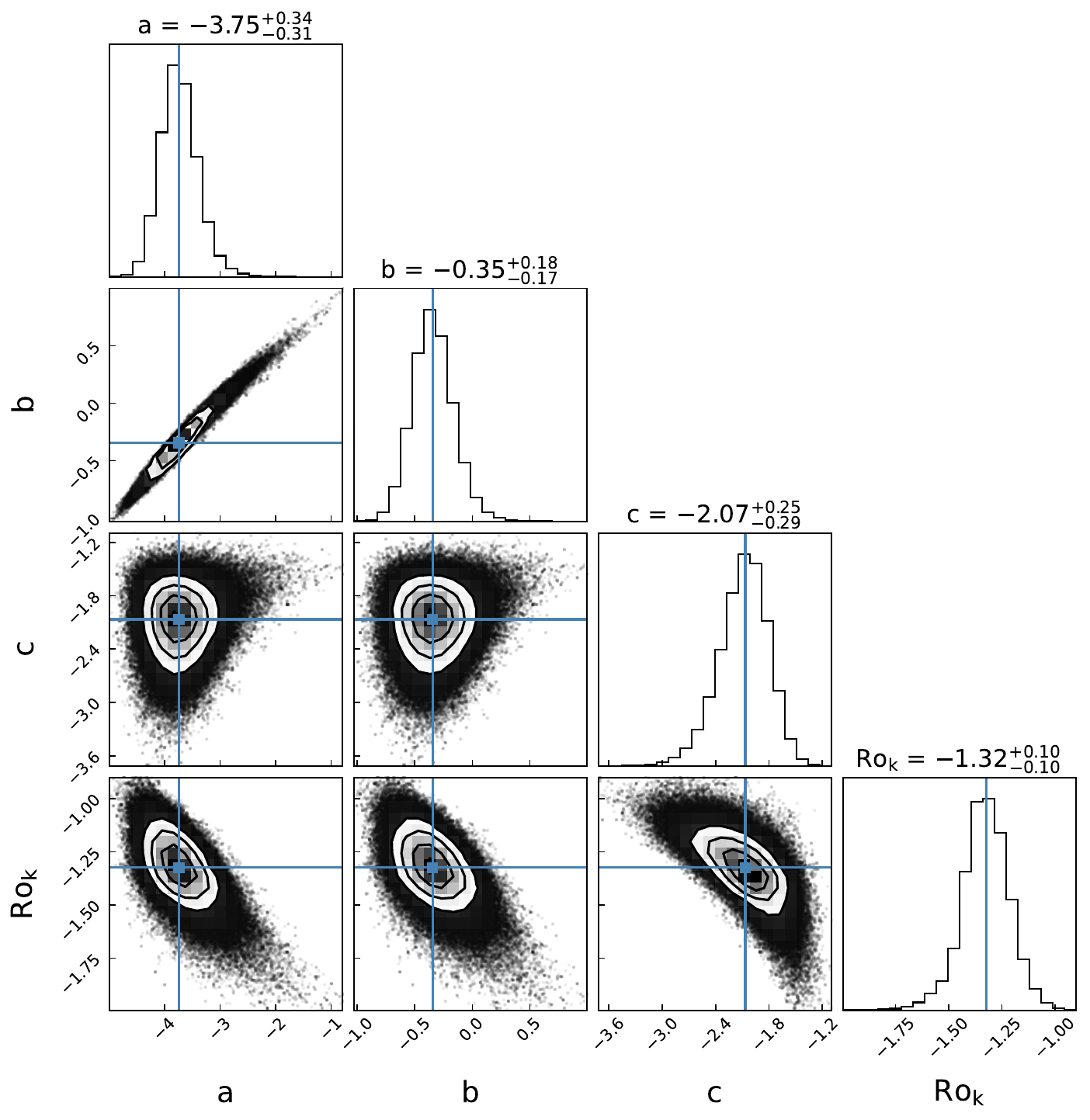}}
%\label{$R_X$.fig}
\caption{Panel (a): posterior distributions of parameters of the piece-wise power law for single stars. Panel (b): posterior distributions of parameters of the piece-wise power law for binaries}. %These figures were plotted based on the \emph{corner} python package from \cite{corner}.} 
%These figures were plotted based on the \emph{corner} python package from \cite{corner}.}
\label{MCMC}
\end{figure*}

\end{appendix}

\end{CJK*}
\end{document}